\documentclass[fleqn,usenatbib]{mnras}
\usepackage{newtxtext,newtxmath}
\usepackage[T1]{fontenc}
\DeclareRobustCommand{\VAN}[3]{#2}
\let\VANthebibliography\thebibliography
\def\thebibliography{\DeclareRobustCommand{\VAN}[3]{##3}\VANthebibliography}

\usepackage{graphicx}	
\usepackage{amsmath}	
\usepackage{physics}

\title[Beta~Pic dust clump motion]{Has the dust clump in the debris disk of Beta~Pictoris moved?}

\author[Han et al.]{
Y.~Han,$^{1}$\thanks{E-mail: yinuo.han@ast.cam.ac.uk}
M.~C.~Wyatt,$^{1}$
W.~R.~F.~Dent,$^{2}$
\\
$^{1}$Institute of Astronomy, University of Cambridge, Madingley Road, Cambridge CB3 0HA, UK\\
$^{2}$ALMA JAO, Alonso de Cordova 3107, Casilla 763 0355, Santiago, Chile
}

\date{Accepted XXX. Received YYY; in original form ZZZ}

\pubyear{2022}

\begin{document}
\label{firstpage}
\pagerange{\pageref{firstpage}--\pageref{lastpage}}
\maketitle

\begin{abstract}
The edge-on debris disk of the nearby young star Beta~Pictoris shows an unusual brightness asymmetry in the form of a clump. The clump has been detected in both the mid-IR and CO and its origin has so far remained uncertain. Here we present new mid-IR observations of Beta~Pic to track any motion of the dust clump. Together with previous observations, the data span a period of 12 years. We measured any projected displacement of the dust clump over the 12-yr period to be $0.2^{+1.3}_{-1.4}$~au away from the star based on the median and $1\sigma$ uncertainty, and constrain this displacement to be $<$11~au at the $3\sigma$ level. 
This implies that the observed motion is incompatible with Keplerian motion at the 2.8$\sigma$ level. It has been posited that a planet migrating outwards may trap planetesimals into a 2:1 resonance, resulting in the observed clump at pericentre of their orbits that trails the planet. The observed motion is also incompatible with such resonant motion at the 2.6$\sigma$ level. 
While Keplerian motion and resonant motion is still possible, the data suggest that the dust clump is more likely stationary. Such a stationary dust clump could originate from the collision or tidal disruption of a planet--sized body, or from secular perturbations due to a planet that create regions with enhanced densities in the disk. 
\end{abstract}

\begin{keywords}
circumstellar matter -- stars: individual: Beta~Pic -- planet--disc interactions
\end{keywords}


\section{Introduction}
Circumstellar debris disks are ubiquitous in mature planetary systems. The growing sample of recent observations have demonstrated that debris disks are rich in substructures \citep{Hughes2018}. These structures may be linked to the dynamical history within the system and be used to probe embedded bodies such as planets that may otherwise be hidden from observation. 

The nearby (19.6~pc, \citealp{Gaia2018}) young A5V \citep{Houk1978} star Beta~Pictoris hosts a bright edge-on debris disk that is among the most extensively studied. At an age of 21~Myr \citep{Binks2014}, its debris disk is one of the youngest known at such proximity to Earth. Observations of Beta~Pic have revealed a host of interesting substructures across different wavelengths. In the scattered light, the disk exhibits a length asymmetry with a longer NE arm \citep{Larwood2001} and a secondary disk that is tilted relative to the primary disk \citep{Golimowski2006, Apai2015}. The resulting warp-like geometry has been linked to a perturbing planet \citep{Lagrange2009}. At mm wavelengths, the vertical structure of the disk has been suggested to be consistent with a dynamically hot and cold population of planetesimals, analogous to the the Solar System’s Kuiper Belt \citep{Matra2019}. 

However, one particularly prominent feature of the disk whose interpretation remains unresolved is the presence of a clump on the SW arm, which was imaged in the mid-infrared (IR) by \citet{Telesco2005}, following an earlier detection of an asymmetry between the two arms \citep{Pantin1997}. Follow-up observations by \citet{Li2012} using the same instrument marginally suggested a projected outwards displacement of the clump over a 7-yr time baseline. Subsequent ALMA observations \citep{Dent2014, Matra2017} also detected a co-located clump in CO. CO observations carry velocity information. Assuming that the gas undergoes Keplerian orbital motion, the CO observation points to the clump being at a radius of 85~au from the star when deprojected  \citep{Dent2014}. 


The existence of clumps has previously been suggested in other debris disks and has been linked to dynamical interactions with planets \citep{Ozernoy2000, Wilner2002, Quillen2002}. For example, dust clumps have been suggested to exist in Epsilon Eridani, and their orbital motion was tracked with observations at mm wavelengths spanning a 5-yr time baseline \citep{Poulton2006}. However, follow-up observations of Epsilon Eridani \citep{Chavez2016} and other systems previously thought to be clumpy such as Vega \citep{Matra2020} and Fomalhaut \citep{MacGregor2017Fom} have since then suggested that these disks are likely smoother than previously thought, leaving Beta~Pic as the only debris disk known to exhibit such a significant azimuthal structure \citep{Hughes2018}. 

Theories proposed to explain the origin of the mid-IR and CO clump can be broadly classified according to whether or not the clump is expected to be moving over orbital timescales. In the category of a moving dust clump, one theory is that the dust clump originated from a very recent collision, immediately after which it undergoes Keplerian motion \citep{Telesco2005}.  It has also been proposed that the enhanced emission in the clump could be due to dust and gas originating from the collisional grinding of planetesimals trapped in resonance with a planet in the system \citep{Telesco2005, Dent2014}. These resonances create clumps that orbit the star at a period equal to that of the planet, and the precise geometry of the clump depends on the relative fractions of planetesimals in each different type of resonance \citep{Wyatt2003}. Alternatively, the destruction of a planetesimal could set off a collisional avalanche which propagates outwards in the disk and creates regions more abundant in dust \citep{Grigorieva2007}. 

In the category of a stationary dust clump, one scenario is that the clump originates from the break--up of a large body via either a giant impact \citep{Telesco2005, Dent2014, Jackson2014} or a tidal disruption event \citep{Cataldi2018}. The orbits of the resulting debris vary, but they pass through the site of collision, resulting in further collisions and enhanced emission in that region. Although the probability of witnessing such an event immediately after its occurrence is low, \citet{Jackson2014} showed that such a level of enhanced emission might be possible long after the break--up event if the body was sufficiently large. 

Other models that are consistent with a stationary dust clump include secular perturbations due to an inner planet which could cause material to come together at localised regions of the disk \citep{Nesvold2015}, and an eccentric disk which could result in a brightness asymmetry \citep{Cataldi2018}.

Discerning between the two classes of theories can therefore be achieved with a measurement of the motion of the dust clump. To verify the displacement of the dust clump measurement by \citet{Li2012} and obtain more robust constraints on any proper motion, we observed Beta~Pic in the mid-IR with an additional epoch using VLT/VISIR to compare with previous observations. We describe the new mid-IR observations of Beta~Pic in Section~\ref{sec:obs} and the data reduction procedures in Section~\ref{sec:reduction}. We then constrain the clump motion in Section~\ref{sec:motion} and discuss potential interpretations in Section~\ref{sec:discussion}. Our findings are summarised in Section~\ref{sec:conclusions}.

\section{Observations}
\label{sec:obs}
We observed Beta~Pic with the VISIR camera at the Very Large Telescope (VLT) at Paranal Observatory on two nights in Sep 2015. Beta~Pic has previously been observed on two occasions using the T-ReCS camera on the Gemini South Telescope, and a primary science goal of this observation is to constrain the proper motion of the dust clump in the disk by comparing between these and our new epoch of observation. We therefore used the B11.7 filter on VISIR, which has the most consistent wavelength coverage with the Si5 filter used in previous T-ReCS observations. To characterise the point-spread function (PSF) of this observation, we observed HD50310 as a calibrator star, consistent with the calibrator star in the first epoch of T-ReCS observation.

An observing log summarising all observations used in this study is presented in Table~\ref{obslog}. Within each epoch, we grouped integrations into two observations. The 2015 epoch was observed on two separate nights, which are summarised as two individual observations in the table. The 2003 epoch was observed on the same night, but the observations were divided into two groups by a $\sim$3~hr gap, with the PSF star observed separately within both groups. These are summarised as two observations in the table. The 2010 epoch was observed continuously on the same night with the PSF observed at the beginning of the sequence. However, the last of the 4 integrations was taken under a different position angle of the telescope, which we summarise as a separate observation (2010 Obs.~2). Since the sequence was completed within a relatively compact window of $\sim$1.5~hr, we consider the PSF observation applicable to both groups. 

\section{Data reduction}
\label{sec:reduction}
Based on results from \citet{Li2012}, any motion of the dust clump is expected to be small and sub-pixel level precision is required to track its location. To enable the highest possible precision, we reduced data from all observations (including re-reducing those previously published) to ensure consistency in data reduction. 

\begin{table*}
    \centering
    \caption{Observing log of all Beta~Pic observations used in this study. The new VISIR observations expand on the two previous observations already summarised in \citet{Li2012}.}
    \label{obslog}
\begin{tabular}{lcccccc}
\hline \hline
                                     & \multicolumn{2}{c}{2003}     & \multicolumn{2}{c}{2010}     & \multicolumn{2}{c}{2015}      \\ 
                                     & Obs.~1        & Obs.~2       & Obs.~1        & Obs.~2       & Obs.~1         & Obs.~2       \\ \hline
Date of observation (UT)             & 30 Dec        & 30 Dec       & 16 Dec        & 16 Dec       & 1 Sep          & 15 Sep       \\
Instrument                           & \multicolumn{4}{c}{Gemini/T-ReCS}                           & \multicolumn{2}{c}{VLT/VISIR} \\
Filter name                          & \multicolumn{4}{c}{Si5-11.7um}                              & \multicolumn{2}{c}{B11.7}     \\
Central wavelength / FWHM (microns)  & \multicolumn{4}{c}{11.66 / 1.13}                            & \multicolumn{2}{c}{11.51 / 0.90} \\
Plate scale (arcsec/pixel)           & \multicolumn{4}{c}{0.09}                                    & \multicolumn{2}{c}{0.045}     \\
Integration time (s)                 & 2$\times$478  & 3$\times$456 & 3$\times$319  & 1$\times$203 & 1794            & 1794        \\
Instrument P.A. (deg)                & 340           & 340          & 0             & 32           & -60             & -60         \\
PSF reference star                   & \multicolumn{2}{c}{HD 50310} & \multicolumn{2}{c}{HD 39523} & \multicolumn{2}{c}{HD 50310}  \\
PSF FWHM (arcsec)                    & 0.40          & 0.36         & 0.41          & 0.41         & 0.36            & 0.39        \\ \hline
\end{tabular}
\end{table*}

\subsection{Chop-nod subtraction}
All T-ReCS and VISIR observations used a standard chop-nod cycle, in which the telescope nods between two positions (A and B), and chops between two positions (1 and 2) multiple times at a given nodding position. Following standard mid-infrared chop-nod subtraction routine, within each nodding location in a cycle (e.g., nod A), we summed all frames taken at the same chopping location to obtain a ``stacked chopping frame'' for each chopping location (e.g., chop A1 and A2). We took the difference between the two stacked chopping frames to obtain the ``nodding frame'' (e.g., A = A1 - A2). 

\subsubsection{T-ReCS}
For T-ReCS observations, the chopping distance was large enough such that the target is only present in the frame at one chopping location, resulting in only one copy of the target within each nodding frame. We found that there is considerable random offset in the target's location between different nodding frames taken at the same nodding position (e.g., first nod A frame and second nod A frame). The direction and distance for nodding and chopping are the same, and so the location of the target in neighbouring chopping frames within the cycle (e.g., first nod A frame and first nod B frame) overlap almost completely, but their location also exhibits random offset. To afford the sub-pixel precision required for subsequent proper motion analyses, we re-centred each nodding frame independently before stacking them to obtain a final image for the chop-nod cycle (final image = all nod A frames - all nod B frames). The nodding frames were centred by fitting a 2D Gaussian model to the image and using the peak of the fitted Gaussian as the centre of the image, which is possible to do as half of the emission originates from within a distance of 1 PSF FWHM from the star.

The T-ReCS frames exhibited significant offset from a median background of 0 (i.e., ``DC offset'') in rows and columns of pixels that pass through the PSF core of bright targets. To correct for these read-out artefacts, we first masked the disk emission to find the median value of each row and column in the remaining image of the background. We then subtracted from each row and column of pixels in the original image the median background value of the corresponding row or column.

\subsubsection{VISIR}
For VISIR observations, the chopping distance was small enough such that the sub-images from both chopping locations are visible in the nodding frame, one of which is a positive image and the other negative. Furthermore, the nodding direction is perpendicular to the chopping direction. Subtracting neighbouring nodding frames (e.g., first nod A frame - first nod B frame, etc.) gives a ``chop-nodded frame'' which contains four sub-images, two of which are positive and the other two negative. As before, to achieve the highest possible proper motion precision, we centred each sub-image individually before stacking all sub-images from all chop-nodded frames to obtain the final image (final image = all positive sub-images - all negative sub-images). We centred each sub-image in a given chop-nodded frame by masking out the three other sub-images present in the frame (the approximate locations of which are fixed) and fitting a 2D Gaussian model to find the peak of the emission. 

\subsection{Flux calibration}
We calibrated the flux in our observations against the in-band standard star fluxes provided by the Gemini\footnote{\href{https://webarchive.gemini.edu/20210512-sciops--instruments--michelle/find-band-mid-ir-standard-star-fluxes.html}{https://webarchive.gemini.edu/20210512-sciops--instruments--michelle/find-band-mid-ir-standard-star-fluxes.html}} and Paranal\footnote{\href{https://www.eso.org/sci/facilities/paranal/instruments/visir/tools.html}{https://www.eso.org/sci/facilities/paranal/instruments/visir/tools.html}} Observatories. As we observed the standard stars listed in Table~\ref{obslog} on each night for which Beta~Pic was observed, we used the in-band flux of the corresponding standard star to calibrate the Jy/pixel for that night. 

To measure the total in-band flux of Beta~Pic in each observation, we summed up the flux within a rectangular aperture centred on the star that is 13.5$^{\prime\prime}$ ($\sim$260~au) along the major axis of the disk and 4.5$^{\prime\prime}$ ($\sim$87~au) perpendicular to the major axis of the disk. Here and throughout this paper, all linear on-sky distances assume that the distance of Beta Pic from the Earth is 19.6~pc \citep{Gaia2018}. The calibrated in-band fluxes for all observations used in this study are presented in Table~\ref{angle}. 

We note that the uncertainties in these aperture photometry measurements are derived purely from the background noise and do not take into account the systematic uncertainties of the instruments, which are expected to be significantly larger. For example, VISIR observations of standard stars may be subject to photometric uncertainties of $\sim$3\% even with 3~hours of integration time \citep{deWit2020}. Taking into account the systematic uncertainties of these observations, the aperture photometry across the observations are therefore consistent with each other.

\begin{table*}
    \centering
    \caption{Total calibrated flux of Beta~Pic and the derived position angle of the debris disk from all observations used in this study. Note that the position angle may be subject to an additional uncertainty due to that of the orientation of the detector that is not accounted for in this table. }
    \label{angle}
\begin{tabular}{lcccccc}
\hline \hline
                             & \multicolumn{2}{c}{2003}    & \multicolumn{2}{c}{2010}      & \multicolumn{2}{c}{2015}      \\
                             & Obs.~1      & Obs.~2      & Obs.~1      & Obs.~2        & Obs.~1      & Obs.~2        \\ \hline
Calibrated in-band flux (Jy) & 2.728 $\pm$ 0.003 & 2.726 $\pm$ 0.003 & 2.805 $\pm$ 0.004 & 2.843 $\pm$ 0.008 & 2.815 $\pm$ 0.004 & 2.852 $\pm$ 0.004 \\
Disk position angle (deg)    & 32.5 $\pm$ 0.7   & 32.5 $\pm$ 0.8   & 32.6 $\pm$ 0.9   & 30.8 $\pm$ 1.0   & 31.4 $\pm$ 0.9   & 32.4 $\pm$ 0.5 \\ \hline
\end{tabular}
\end{table*}

\subsection{Position angle of the disk}
We measured the position angle of the disk across all six nights of observations independently. 

To determine the position angle of the disk for a given epoch, we first removed the central stellar emission with a circular mask of radius $r$, which was set to 26~au for T-ReCS images and 17~au for VISIR images. We then defined a rectangular region that is 350~au wide in the East-West direction, $h$~au wide in the North-South direction and centered on the star. We then rotated the image by an angle, $\theta$, and calculated the total flux that falls within the rectangular region, $f$. We determined the position angle using the value of $\theta$ that maximises $f$. 

Since the position angle determined in this way varies slightly depending on the choice of $h$, we repeated this procedure for a range of different values of $h$ for each epoch of observation. We chose the values of $h$ to be evenly spaced between 17 and 66~au with a step size of 3.5~au (two T-ReCS pixels, or four VISIR pixels). For any value of $h$ lower than this range, the value of $f$ is heavily affected by noise, whereas for any value larger than this range, the height of the rectangle becomes significantly larger than the height of the disk emission. We used the mean and standard deviation among all fitted position angles to estimate the best-fit position angle and its uncertainty. 

\begin{figure*}
    \centering
    \includegraphics[width=18cm]{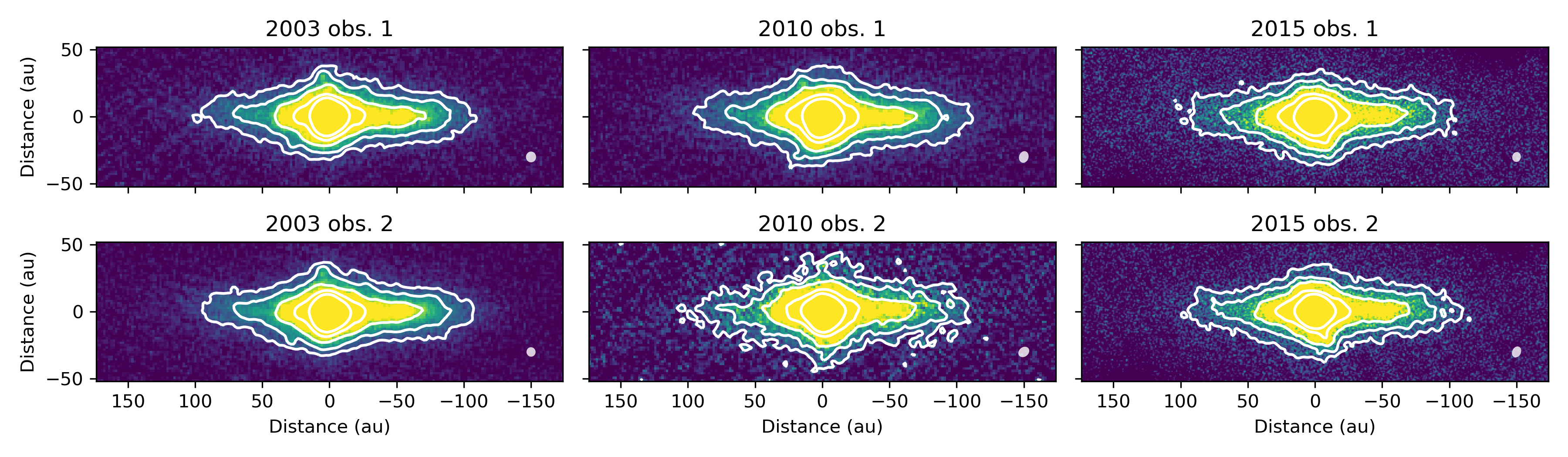}
    \caption{This figure displays the reduced images from all Beta~Pic observations used in this study. The images are rotated according to their respective fitted position angles (Table~\ref{angle}) and are stretched between 0 and 0.062~Jy/arcsec$^2$ for display. The SW arm is to the right of the star. The contours are calculated using the reduced images convolved with a Gaussian kernel with a standard deviation of 0.09~arcsec (1.7~au, or 1 T-ReCS pixel or 2 VISIR pixels) to mitigate the influence of noise, and are drawn at 0.012, 0.025, 0.049, 0.099 and 0.198~Jy/arcsec$^2$ respectively. The ellipse in the bottom-right corner of each panel indicate the PSF FWHM, which is measured by fitting a 2D Gaussian model to each PSF star. }
    \label{fig:images}
\end{figure*}

The fitted position angles (counterclockwise relative to North) are presented in Table~\ref{angle} and the P.A.-rotated images are presented in Fig.~\ref{fig:images}. 

The mean mid-IR position angle across all observations is $32.0 \pm 0.8^\circ$. This value is broadly consistent with the position angle of 33$^\circ$ found by \citet{Telesco2005} based on only the 2003 observations. In the near-IR, \citet{Apai2015} measured a position angle of $29.1 \pm 0.1^\circ$ using the position angle of the brightest pixel at approximately 200~au from the star in both arms. This near-IR position angle is based on emission from a significantly larger distance from the star, and is smaller than the position angle found in this study. In these scattered light images, a warp-like structure (or tilted secondary disk) has been observed in both arms, rising above the midplane on the SW arm and below in the NE arm. The position angle discrepancy could imply that the mid-IR emission at this distance is also affected by the warp observed in the scattered light, the tilt of which could act to increase the apparent position angle within $\sim$100~au.

\section{Has the dust clump moved?}
\label{sec:motion}
In this section, we measure the proper motion of the dust clump using all observations presented in Sec.~\ref{sec:reduction}. We discuss constraints that our measurements set on the presence, orbit and migration history of any potentially perturbing planets in the system under models of the origin of the dust clump in Section~\ref{sec:discussion}. 

\subsection{Isolating the dust clump}
\label{sec:isolate}
\begin{figure*}
    \centering
    \includegraphics[width=18cm]{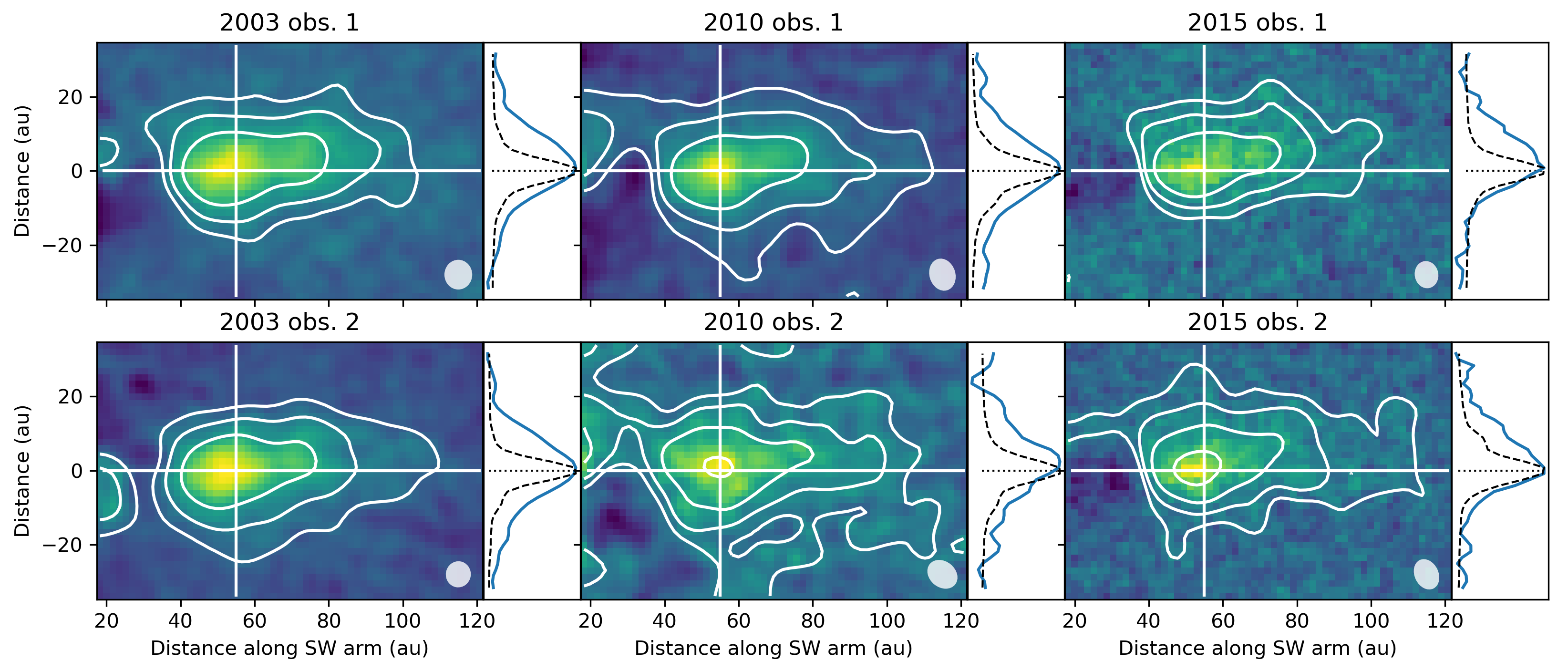}
    \caption{Images of the dust clump isolated from each observation by subtracting the emission from the opposite side of the disk. The contours are calculated using the clump images convolved with a Gaussian kernel with a standard deviation of 0.09~arcsec (1.7~au, or 1 T-ReCS pixel or 2 VISIR pixels) to mitigate the influence of noise, and are drawn at at 0.005, 0.010, 0.019 and 0.037 Jy/arcsec$^2$. The FWHM of the PSF is indicated in the bottom-right corner of each panel. The panel to the right of each clump image shows the normalised vertical flux profile of the clump integrated from 35~au to 104~au along the SW arm. The normalised vertically integrated flux profile of the PSF is shown with a dashed line.}
    \label{fig:clump}
\end{figure*}

We assumed that the mid-IR disk emission consists of two components: an axisymmetric disk plus a dust clump on the SW arm (Fig.~\ref{fig:images}). Under this assumption, we isolated the dust clump by rotating each image by 180 degrees about the centre of the star and subtracting the rotated image from the original image. This is equivalent to subtracting off emission from the opposite location in the disk, effectively removing the underlying disk and leaving only the dust clump. Images of the isolated dust clump and non-rotationally symmetric emission are presented in Fig.~\ref{fig:clump}. 

Fig.~\ref{fig:clump} also shows the horizontally integrated flux of each clump image, which is observed to be slightly offset above the midplane. The scattered light warp, which has been suggested to be a result of perturbations from the planet Beta~Pic~b \citep{Lagrange2009}, is observed to be at the same projected distance as the dust clump \citep{Golimowski2006, Apai2015}. The vertical offset of the clump is in the same direction as the offset of disk emission created by the warp at this location, lending further credence to the idea that the two structures could be related \citep{Nesvold2015}.

\subsection{Approach for constraining displacement}
The dust clump is well resolved and shows a consistent geometric structure across all observations, enabling comparison of the location of the clump. However, each image corresponds to a different underlying PSF, and therefore a direct comparison across all epochs does not afford the highest accuracy required for our analysis of very small displacements. 

To directly compare between any two images, they must first be convolved with each other's PSF. Given the set of 6 observations, there are 15 possible combinations of pairwise comparisons. Our approach is to treat each image as an independent observation (including two images observed in the same year), defining a method (Sec.~\ref{sec:pairwise}) to measure their relative locations. We then solve for the optimal configuration that best gives rise to the 15 relative displacements observed, effectively placing the locations of the 6 dust clump images on a 1D axis along the disk's major axis. Finally, we fit a model to the 6 sample points to constrain the amount of underlying displacement over all observations.

\subsection{Measuring the displacement between two observations}
\label{sec:pairwise}
To measure the relative displacements of the dust clump between any two images indexed $i$ and $j$, we first convolved the full images with each other's PSF:
\begin{equation}
    I_{i, \text{conv-}j}(x, y) = I_{i}(x, y) \ast \text{PSF}_{j}(x, y),
\end{equation}
\noindent where $I_i$ is the image of observation $i$, $\text{PSF}_j$ is the PSF of the observation $j$ and $x$ and $y$ are coordinates in the direction along and perpendicular to the disk's major axis respectively. We then subtracted the opposite side of the disk to isolate the dust clump using the procedure described in Sec.~\ref{sec:isolate}. 

Note that this procedure is different from first subtracting the opposite side of the disk and then convolving with the other PSF (i.e., directly convolving the images in Fig.~\ref{fig:clump} with other PSFs). In the former case, any asymmetries of the PSF of one image also become reproduced in the other, whereas in the latter case they are not. The former procedure therefore allows for a higher degree of consistency when comparing two images, and is expected to mitigate effects such as the apparent peak of the dust clump being slightly shifted due to changes in the PSF between datasets. Appendix~\ref{appendix:subtraction} illustrates the benefits of the former procedure in more detail. 

Since our primary goal is to determine any horizontal displacement of the dust clump, to improve the S/N before making any measurements, we integrated the flux vertically to reduce the clump image into a 1D observable, $J(x)$:
\begin{equation}
    J_{i, \text{conv-}j}(x) = \int_{-h}^{h} I_{i, \text{conv-}j}(x, y) dy,
\end{equation}

\noindent where $h$ was set to 17~au. 

Asymmetries in the PSF mean that a central point source will not be perfectly subtracted in the rotational subtraction process. The bright core centred on the star is resolved in the images and is significantly brighter than the disk. To ensure that any PSF asymmetries associated with the bright core do not extend up to the location of the dust clump, we modelled the core emission with a circular disk with a diameter of 5~au and convolved it with all pairs of PSFs among the observations. Rotationally subtracting the doubly-convolved core model shows that in the worst case among all pairs of PSFs, the effects due to asymmetry of the bright core amounts to 17\% of the peak flux profile of the dust clump at worst at any horizontal location beyond 35~au, and 9\% beyond 40~au, which is approximately the location of the inner edge of the dust clump.
The emission from the dust clump isolated from the rotational subtraction process can therefore be considered largely free from any significant effects of the bright core's asymmetry. Any emission after subtraction that is within 35~au should be treated with caution since it could originate from asymmetries in the PSF.

\begin{figure}
    \centering
    \includegraphics[width=8cm]{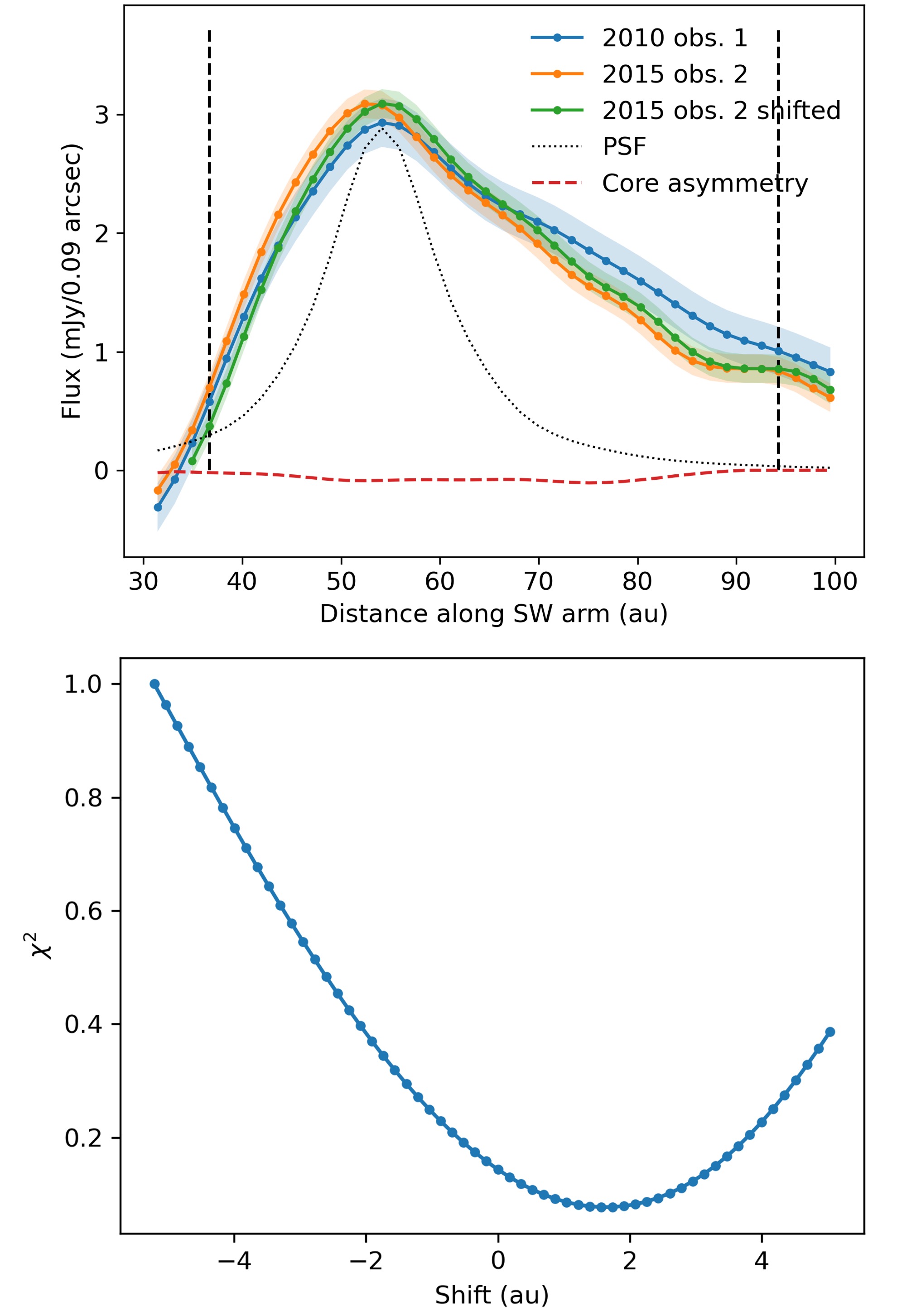}
    \caption{Top panel: the vertically integrated flux of the clump from 2010 Obs.~1 (blue), 2015 Obs.~2 (orange) and 2015 Obs.~2 shifted by an amount (green) that minimises the squared difference compared to 2010 Obs.~1 within the region between the two vertical dashed lines. As the two images have been convolved with each other's PSF, the PSF underlying the two curves is effectively the two original PSFs convolved with each other, the vertically integrated and vertically scaled flux profile of which is shown with the dotted curve. The dashed red curve shows the effect of PSF asymmetry due to the bright core, which is estimated based on the aforementioned effective PSF and a simple model approximating the core as a circular disk. The effect of PSF asymmetry is small in the range displayed, but becomes significant within 30~au. Bottom panel: the squared difference ($\chi^2$) between the two vertically integrated fluxes as a function of how much the 2015 Obs.~2 curve is shifted. The $\chi^2$ values are normalised to the maximum values in the plot. }
    \label{fig:shift}
\end{figure}

The vertically summed flux as a function of distance, $x$, has three main features: an inner edge, a peak and an outer edge (see Fig.~\ref{fig:shift}). Given the consistency in the geometry of the clump, we define the relative displacement between the two curves as the amount of horizontal shift required of one curve to produce maximal overlap with the other. Specifically, we determine this by fixing one curve and shifting the other by 0.17~au (or 0.1 Gemini pixels) at a time, computing the squared difference between the two curves at each new location:
\begin{equation}
    \chi^2_{ij}(\delta) = \int_{x_1}^{x_2} (J_{i, \text{conv-}j}(x) - J_{j, \text{conv-}i}(x - \delta))^2 dx,
\end{equation}
\noindent where $x_1$ was set to 36~au and $x_2$ to 96~au to cover the extent of the clump. The values of $x_1$ and $x_2$ were determined using the mean inner and outer bounds for which $J_{i, \text{conv-}j}$ is above a quarter of its peak value. The relative displacement between the two clumps is then defined as the amount of shift, $\delta_{ij}$, which minimises the $\chi^2_{ij}$. 

An advantage of this method over other definitions of clump location, such as using only the peak location, is that all three features of the clump are simultaneously used to determine its relative position, further mitigating the impact of noise. An example plot of the squared difference between two curves as a function of the amount of shift is shown in Fig.~\ref{fig:shift}. The smooth variation of the curve defines a clear minimum. The optimal shift as determined by this method is plotted in Fig.~\ref{fig:shift} and is a reasonable estimate by inspection, neatly aligning the peak and edges of the two curves.

We performed the analysis on all 15 combinations of observation pairs. The results may be conveniently summarised into a ``relative displacement matrix'', $\Delta_\text{obs}$:
\begin{equation}
    \Delta_{\text{obs, }ij} = \begin{cases}
    \delta_{ij} \ , \ i \ne j,\\
    0 \ , \ \text{otherwise}.
    \end{cases}
\end{equation}

\noindent The matrix is shown graphically in Fig.~\ref{fig:matrix}. 

\subsection{Finding a consistent spatial configuration}
With 15 combinations of pairwise relative displacements, we devised a spatial configuration of the 6 clumps that best satisfies all observed relative positions. 

We constructed the problem by fixing the location of the clump in 2015 Obs. 2 at 0, and defining the coordinates of all other clump locations relative to this observation, giving rise to 5 free parameters. We also assumed independently and identically distributed Gaussian errors across all entries in $\Delta_{\text{obs}}$, with the standard deviation defining the errors left as a free parameter. 

\begin{figure}
    \centering
    \includegraphics[width=8cm]{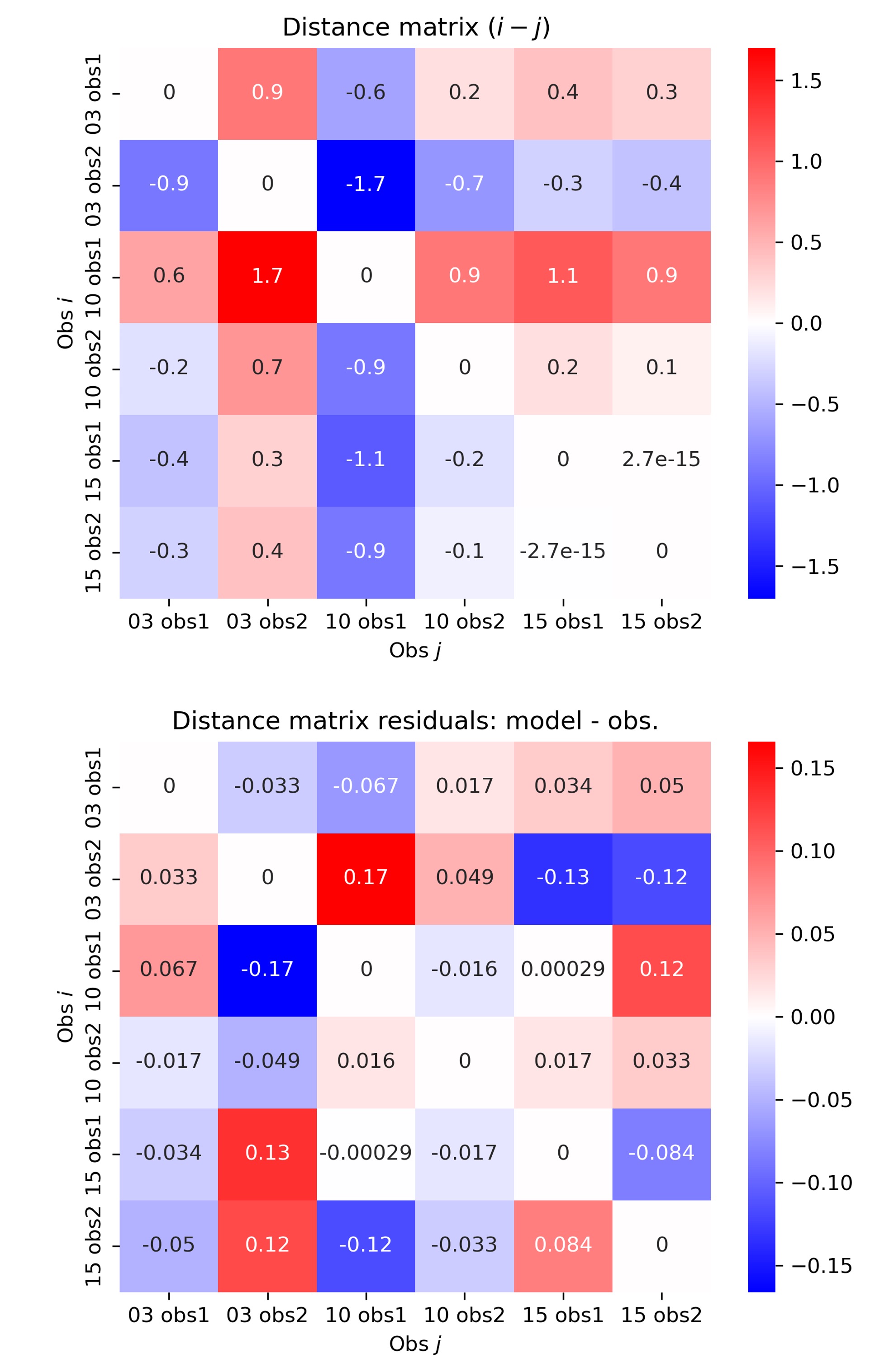}
    \caption{Top panel: relative displacement matrix of the observations, which contains all pairs of relative displacements measured. Entries in the leading diagonal are always zero. A significant and consistent displacement would be represented by a matrix in which values become increasingly positive (red) towards the lower-left and increasingly negative (blue) towards the upper-right. Bottom panel: residuals of the displacement matrix calculated from the best-fit model of the spatial configuration of all clump observations. }
    \label{fig:matrix}
\end{figure}

We used a Markov Chain Monte Carlo (MCMC) approach to sample the 6-dimensional parameter space implemented with the \texttt{emcee} package \citep{emcee}. At each step of the MCMC, the relative displacement matrix of the proposed spatial configuration, $\Delta_{\text{mod}}$, was computed and the log-likelihood was computed against $\Delta_{\text{obs}}$. Such a method effectively finds the most likely spatial configuration and its uncertainties that best reproduces all observed pairwise displacements. We estimated the best-fit location of each clump observation relative to 2015 Obs. 2 using the median of the marginalised posterior distributions. The results are presented in Table~\ref{configuration} and the residual relative displacement matrix of the best-fit model is shown in Fig.~\ref{fig:matrix}. 

\begin{table}
    \centering
    \caption{Fitted locations of the dust clump in each observation relative to 2015 Obs.~2 and the fitted 1$\sigma$ uncertainties on pairwise distances $\delta_{ij}$. }
    \label{configuration}
    \begin{tabular}{c|c}
        \hline \hline
        Observation & Value (au) \\
        \hline
        2003 Obs.~1 & 0.61 $\pm$ 0.06 \\
        2003 Obs.~2 & -0.90 $\pm$ 0.06 \\
        2010 Obs.~1 & 1.77 $\pm$ 0.06 \\
        2010 Obs.~2 & 0.23 $\pm$ 0.06 \\
        2015 Obs.~1 & -0.15 $\pm$ 0.06 \\
        2015 Obs.~2 & 0 (fixed) \\
        \hline
        1$\sigma$ uncertainties on $\delta_{ij}$ & 0.10 $\pm$ 0.01 \\
        \hline
    \end{tabular}
\end{table}

We highlight that the uncertainties estimated using the MCMC only addresses the requirement that there exists a self-consistent spatial configuration of all observations that is able to explain the observed pairwise relative displacements. It does not enforce the requirement that two observations from the same year should be at approximately the same location. In the following section, we enforce this condition by fitting a linear model through the locations of the 6 clumps as a function of the time of observation and re-estimate the uncertainties upon asserting this linear relationship. This relies on the fact that there are two sets of observations at each epoch, which provides more realistic estimates on each location measurement.

\subsection{Fitting a linear displacement model}
Given the deprojected radial distance of the clump detected in CO, any Keplerian motion should have a period of approximately 600~yr \citep{Dent2014}. Any such motion within the time interval between our observations is therefore expected to be small, and we may approximate the small projected motion of the dust clump over our observations using a linear model. 

\begin{figure}
    \centering
    \includegraphics[width=9cm]{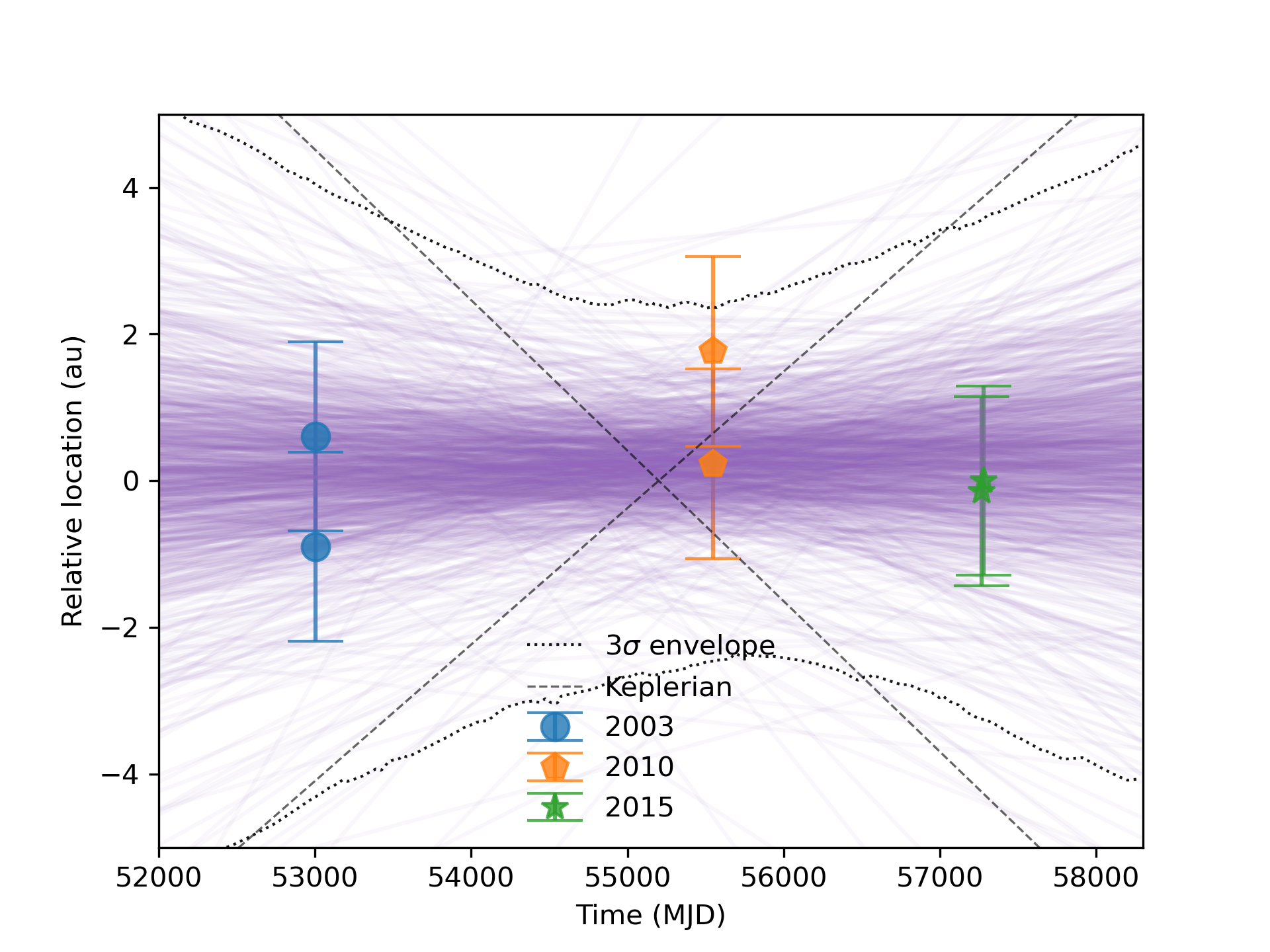}
    \caption{Location of the dust clump relative to 2015 Obs.~2 derived from pairwise displacements with uncertainties derived from the MCMC fit to the linear model. 1000 models randomly drawn from the posterior distribution are over-plotted in purple. The 3$\sigma$ envelope of the models is indicated with dotted lines. The Keplerian orbital speed is indicated with dashed lines. }
    \label{fig:line}
\end{figure}

We fitted a linear model to the 6 epochs of observations using an MCMC implemented with \texttt{emcee} \citep{emcee}. We assumed that the uncertainty on each location determined in the previous section is independently and identically Gaussian distributed, and included the uncertainty as a free parameter in the MCMC. The results of the MCMC are presented in Table~\ref{line}. Fig.~\ref{fig:line} shows 1000 models randomly drawn from the posterior distribution. 

\begin{table*}
    \centering
    \caption{Best-fit linear model of the dust clump motion and the 68\% (1$\sigma$), 90\% (2$\sigma$) and 99.4\% (3$\sigma$) confidence intervals. }
    \label{line}
    \begin{tabular}{l|c|c|c|c}
        \hline \hline
        & Median & 1$\sigma$ & 2$\sigma$ & 3$\sigma$ \\\hline
        Linear displacement speed (au/yr) & 0.018 & (-0.099, 0.13) & (-0.29, 0.33) & (-0.86, 0.90) \\
        Location at MJD = 0 (au) & -2.4 & (-20, 15) & (-49, 45) & (-137, 130) \\
        1$\sigma$ uncertainties on clump location (au) & 1.3 & (0.87, 2.2) & (0.65, 4.3) & (0.50, 11.3) \\
        \hline
    \end{tabular}
\end{table*}

Contrary to the uncertainties obtained in the previous section, the fitted uncertainties on the location of the clump now take into account the systematic uncertainties such as those from the observations and from the methods applied to isolate the clump. For example, although Fig.~\ref{fig:shift} suggests the method comparing the clump location defines a clear optimum, the observed location and shape of the dust clump may have been biased by noise and PSF instabilities. The uncertainties estimated upon asserting that there exists a linear relationship between the location measurements shall therefore be considered as being more realistic. 

Our results suggest that over the 12-yr period spanned by the observations, the projected displacement of the clump along the disk's midplane is constrained to be less than 11~au away from the star and less than 10~au towards the star at the $3\sigma$ level, with a median of 0.2~au away from the star. 
We observe that the two location measurements for the 2015 epoch appear to be more consistent with each other than pairs of observations within the other two epochs are (Fig.~\ref{fig:line}). We conservatively assumed that all epochs share the same uncertainties, however the higher degree of consistency in the 2015 epoch may be reflective of the fact that the images in this epoch achieved a higher sensitivity than in previous epochs. If that is the case, the $3\sigma$ constraints may be even tighter than the more conservative values reported here. 

\section{Discussion}
\label{sec:discussion}
In the following sections, we discuss new constraints that can be obtained in light of the new observations under theories proposed to explain the origin of the dust clump. 

\subsection{Scenarios in which the dust clump is moving}
\label{sec:resonance}
CO velocity data reveals that the clump is located at a radius of 85~au from the star \citep{Dent2014} and that the disk is rotating in a way such that the SW arm is approaching us. However, we do not know whether the dust clump is in front of or behind the line of nodes, resulting in a two-fold degeneracy in spatial configuration of the dust clump relative to the star. The most straight--forward interpretation of the clump is that it orbits the star with Keplerian motion. Assuming a stellar mass of 1.75~$M_\odot$ \citep{Crifo1997}, a Keplerian orbit at 85~au would imply an orbital motion of 0.90~au/yr. Over the time--span of the observations, this would correspond to a mean projected speed of of 0.75~au/yr inwards if the clump is in front of the line of nodes, or 0.68~au/yr outwards if behind. This implies that the observed displacement of the dust clump (or rather the lack of it) is inconsistent with Keplerian motion at the 2.8$\sigma$ level. 

However, the clump's motion is not required to be Keplerian, since one hypothesis for its origin is that it results from collisions among planetesimals trapped in resonance with a perturbing planet. During the formation and evolution of a planetary system, the orbit of a planet can migrate. If a planet migrates outwards, planetesimals that are initially outside its orbit may become trapped in resonance with the planet. The orbits of the resonantly trapped planetesimals are then swept outwards with the outwardly migrating planet while maintaining their resonant period ratio with the planet. Although the planet is assumed to maintain a circular orbit throughout the process, the Keplerian orbits of the resonantly trapped planetesimals increase in both semi-major axis and eccentricity, collectively forming clump-like structures near periastron that orbit the star with the same period as the perturbing planet \citep{Wyatt2003}. However, since the periastra of the orbits of the planetesimals could be different from the orbital radius of the planet, the clumps could therefore appear to undergo non-Keplerian motion. 

\begin{figure}
    \centering
    \includegraphics[width=8cm]{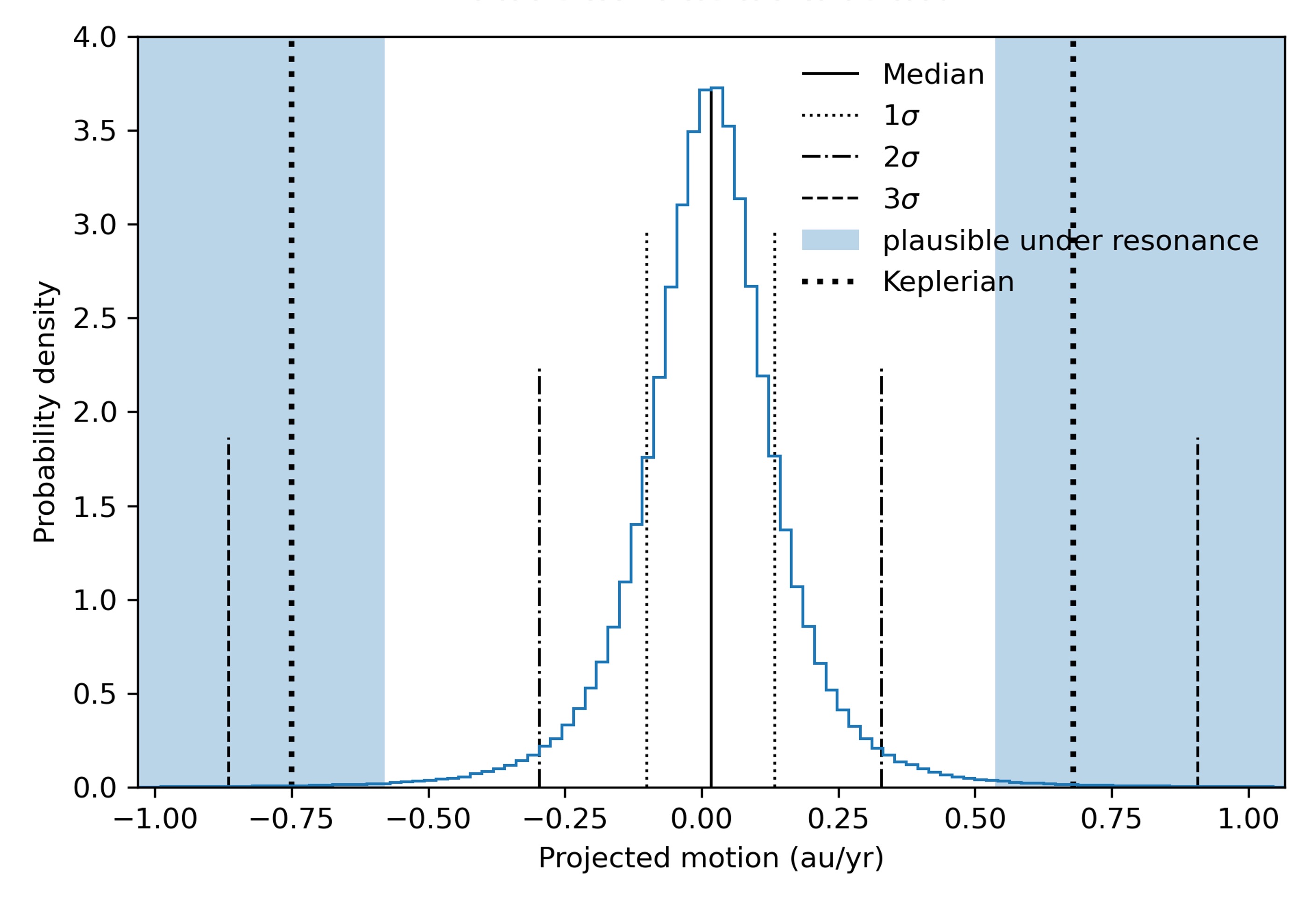}
    \caption{Posterior distribution of the projected motion of the clump over the 12-yr time baseline of our observations. Displacement away from the star is positive. The median, 1$\sigma$, 2$\sigma$ and 3$\sigma$ intervals are labelled with solid, dotted and dash-dotted and dashed lines respectively. Projected motion due to Keplerian motion is labelled with thick dotted lines. The shaded region (which extends beyond the left and right boundaries of this plot) indicates the range of projected motion compatible with the resonance model. The resonance model requires that any perturbing planet should be within 100~au, which corresponds to relatively short orbital periods and therefore fast resonant planetesimal clump motion in this plot. }
    \label{fig:dist}
\end{figure}

Fig.~\ref{fig:dist} shows the distribution of the projected clump motion inferred from the observations, which is equivalent to the distribution of the fitted slopes in Fig.~\ref{fig:line}. Assuming that the planet and dust clump are rotating in the same direction as the disk, a projected clump motion in either direction is possible under the resonance model. However, any migrating planet capable of sweeping planetesimals outwards is not expected to be located outside 100~au where the disk's surface density peaks \citep{Matra2019}. Given the dust clump's 52~au projected distance from the star and 85~au orbital radius, the 100~au upper limit\footnote{Note that although the resonantly trapped planetesimals would technically be on crossing orbits with the perturbing planet if it is outside 85 au, the planetesimals would not collide with the planet since the resonance keeps them separated.} for the planet's semi-major axis means that any projected displacement less than $\sim$~0.55~au/yr in either direction is inconsistent with the resonant planetesimal model. Since only the two tails of the distribution correspond to resonant motion by a Keplerian perturber within 100~au (Fig.~\ref{fig:dist}), the displacement inferred from the data appears to disfavour the resonance model. 

The lack of large observed motion of the dust clump may be consistent with suggestions based on C\textsc{i} observations, which showed its distribution to exhibit a similar asymmetry to CO \citep{Cataldi2018}. C\textsc{i} is a product of the photodissociation of CO and is thought to persist on a timescale longer than the orbital period. It is therefore expected to be spread evenly around the orbit if the dust clump orbits the star. Assuming that no C is removed from the system, \citet{Cataldi2018} proposed that the C\textsc{i} asymmetry could suggest that the asymmetric structure is not in orbit around the star. 

However, as Fig.~\ref{fig:line} suggests, models in which the dust clump is moving, such as the resonance model, cannot be conclusively ruled out. We therefore consider possible scenarios which may plausibly produce the observed dust clump under the resonance model in the following section. 

\subsubsection{Resonance model}
To constrain the orbit of the perturbing planet, we make two major assumptions about the resonant scenario. Firstly, we assume that all planetesimals are trapped in the 2:1(l) resonance \citep{Wyatt2003} in which only one clump is present, as the imagery suggests. 
This assumption implies that the semi-major axes of the planetesimals and planet follow a $2^{2/3}$:1 ratio throughout the migration process upon trapping, and that the clump of planetesimals lags the planet by an angle of approximately 90$^\circ$, with the precise value depending on the eccentricity of the planetesimals. 
Secondly, we assume that the periastron of the resonantly trapped planetesimals ($q$) is equal to the orbital radius of the CO clump at 85~au based on CO velocity data \citep{Dent2014}. 
In reality, the periastron of the orbit of the planetesimals would be slightly interior to the location of the clump, and further modelling would be required to derive its precise value. 

With knowledge of the orbital radius (85~au) and projected separation (52~au) of the dust clump, the projected speed in Fig.~\ref{fig:dist} can be translated into the orbital period of the dust clump, which is equal to the orbital period of the perturbing planet. 
Assuming a stellar mass of 1.75~$M_\odot$, the 3$\sigma$ upper bound on projected speed of $<$0.9~au/yr rules out the presence of any perturbing planet with an orbital radius less than $r_\text{min} = 69$~au. The present orbital radius of the perturbing planet ($r_\text{pl}$), if the dust clump originates from planetesimals in its resonances, is therefore expected to be between 69 and 100~au. 

Assuming that all planetesimals were initially on circular orbits at the same radius, $a_0$, the semi-major axis ($a$) and eccentricity ($e$) of the planetesimals during migration are related by $e^2 = \ln(a/a_0)/2$ (Eq.~22, \citealt{Wyatt2003}).

Since $a = 2^{2/3} r_\text{pl}$ for a 2:1 resonance and $q = a (1 - e)$, the range of $r_\text{pl}$ sets the range of eccentricities, $e = 1 - q/(2^{2/3} r_\text{pl})$, and starting semi-major axes,
\begin{equation}
\label{eq:a0}
    a_0 = 2^{2/3} r_\text{pl} \exp{-2 \bigg( 1 - \frac{q}{2^{2/3} r_\text{pl}} \bigg) }, 
\end{equation}
\noindent implying that the final eccentricities of planetesimals are between 0.22 and 0.46 and the initial semi-major axes between 99 and 103~au before trapping. The range in initial semi-major axes is narrow since a larger orbital radius of the planet corresponds to a higher planetesimal eccentricity, thereby requiring more migration which offsets the larger final semi-major axis. 

The emission of the clump is dominated by material near pericentre, but some of its emission arises from slightly further out, and so the pericentre only defines the smallest possible radial location of a clump. Accounting for the fact that the true periastron of the planetesimals should be slightly less than the assumed value of 85~au to give rise to a clump at 85~au, the true planetesimal eccentricities would be expected to be slightly higher than those presented here. This would in turn imply smaller initial semimajor axes of the planetesimals.

The orbital parameters of the planet and planetesimals under the model set constraints on the migration of the planet. 
We find that a migration of at least 6~au (from 63~au to 69~au) and at most 35~au (from 65~au to 100~au) is required of the planet after resonantly trapping planetesimals.
Since the migration timescale must be less than the age of Beta~Pic, which is estimated to be 21~Myr \citep{Binks2014}, the migration speed must have been greater than 0.3~au/Myr. 

\begin{figure}
    \centering
    \includegraphics[width=9cm]{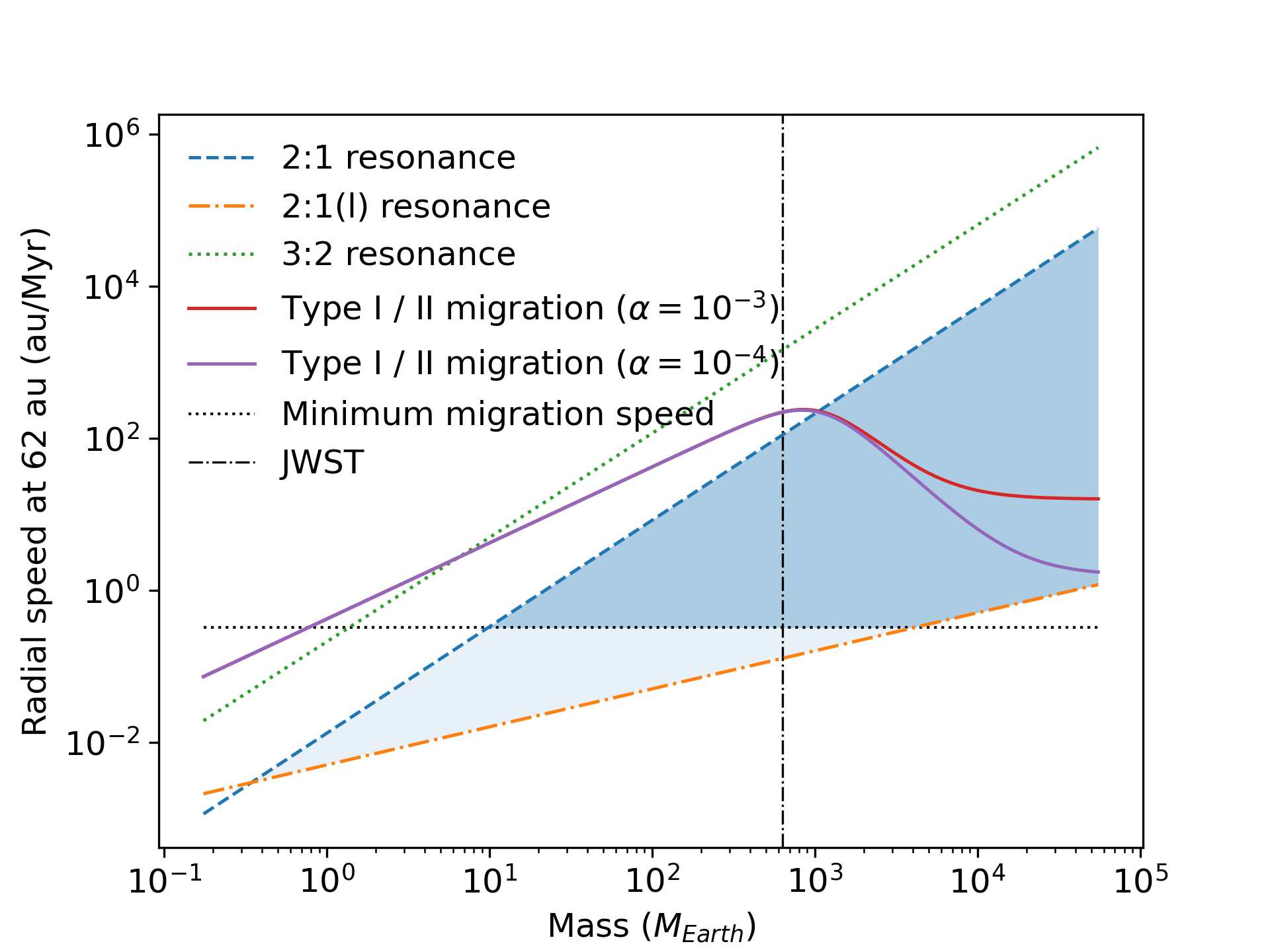}
    \caption{This plot shows, for the Beta Pic system, the 50\% probability thresholds for capturing planetesimals into the 2:1 (dashed) and 3:2 (dotted) resonances during planet migration, and the threshold above which the 2:1(l) resonance is twice as likely as the 2:1(u) resonance (dotted) \citep{Wyatt2003}. The region of parameter space corresponding to the 2:1(l) resonance model discussed in this study is shaded. The horizontal dotted line shows the 3$\sigma$ lower bound for the migration speed at 62~au required by the model inferred from the dust clump's proper motion, which further limits the possible region of parameter space to the darker shaded region. The vertical dash-dotted line shows the minimum planet mass detectable at the 80\% detection threshold with JWST at the range of orbital radii compatible with the perturber required by the resonance model. The solid lines show models \citep{Tanaka2002, Bate2003} of the Type I and Type II gas--driven migration speed at 62~au as a function of planet mass, assuming a disk aspect ratio of 0.1 and a surface density corresponding to the Minimum Mass Solar Nebula \citep{Hayashi1981}. A disk viscosity of $\nu = 10^{-5}$ and $10^{-6}$ times $r^2 \Omega$ (or $\alpha = 10^{-3}$ and $10^{-4}$) is assumed for the two models. }
    \label{fig:migration}
\end{figure}

The probability with which planetesimals are trapped in resonance as a planet migrates depends on both the planet's mass and the radial migration speed relative to the Keplerian orbital speed. Fig.~\ref{fig:migration} provides limits within the mass and migration speed parameter space required to fall into the 2:1(l) resonance regime \citep{Wyatt2003} at 62~au, the lower bound for the the radius at which the planet resonantly traps the planetesimals. Based on these limits and our derived lower bound for the migration speed, the mass of the perturbing planet is required to be at least 10~$M_\text{Earth}$. 

Fig.~\ref{fig:line} shows the gas--driven migration rates expected under hypothetical disk parameters. Details of the derivation of possible migration mechanisms are available in Appendix~\ref{appendix:migration}. Taking into account these constraints on migration rate and planet mass, we find that it may be plausible for a massive planet ($\gtrsim$3~M$_\text{Jupiter}$) migrating due to either interaction with a gas disk or scattering of planetesimals to be responsible for the resonance trapping of planetesimals. 
Alternatively, a smaller planet with a mass as low as M$_\text{Uranus}$ migrating outwards via planetesimal--driven migration could also be responsible for the resonance trapping. 
Assuming that the surface density of the gaseous disk of Beta Pic was at least as high as the Minimum Mass Solar Nebula, the lower--mass planet scenario would be unlikely to occur under gas--driven migration, since the predicted migration speed would be too fast such that the resulting primary resonance could be the 3:2 resonance instead. 



\subsubsection{Collisional avalanche}
\citet{Grigorieva2007} found that a collisional avalanche triggered by the destruction of a planetesimal or comet propagates outwards in a spiral--like pattern, resulting in a two--sided brightness asymmetry if viewed edge--on. This feature could persist for 1,000~yr in a Beta Pic--like disk, with a probability of witnessing such an event of a few percent. 
Future work may wish to explore whether such a mechanism could create the distribution and brightness of the dust and gas clump in Beta Pic.
However, since asymmetries produced by a collisional avalanche could rapidly propagate outwards on the timescale of a few orbits \citep{Thebault2018}, the lack of motion of the dust clump also appears to disfavour this scenario.

\subsection{Scenarios in which the dust clump is stationary}
\subsubsection{Giant impact}
Although the observed displacement of the dust clump (or the lack of it) does not conclusively reject models in which the dust clump is moving, it does suggest that alternative models in which the clump is stationary are favoured. Based on the mass of the dust clump inferred from its mid-infrared emission, \citet{Telesco2005} proposed that a possible origin of the dust clump is the collisional break--up of a large planetesimal approximately 100~km in size within $\sim$50~yr ago. However, the probability of witnessing such an event is extremely low due to the short duration of the dust clump produced, and the clump of material produced immediately after the collision is expected to undergo Keplerian motion, which is disfavoured by the lack of motion found in this study. This scenario is further disfavoured by the much larger mass estimate of the dust clump derived from its mm flux, which implies any parent body should be approximately as massive as Mars \citep{Dent2014}. 

\citet{Jackson2014} showed that the collision of a larger, planet--sized body may allow the asymmetry to be witnessed for $\sim$1~Myr after the collision. Under this model, particles originating from the collision take orbits of varying semi--major axis and eccentricity, but all orbits pass through the site of collision, enhancing the emission in that region. Since the site of collision is fixed, the location of the enhanced emission is also expected to remain stationary. A giant impact of a planet--sized body is therefore a more likely scenario than the collision of a large planetesimal.

Theoretically, it is possible for there to exist a sufficient number of Mars--sized bodies which collide frequently enough to make it probable to witness the aftermath of a collision between them.
In order for it to be probable to witness an asymmetric dust distribution due to such a collision, two conditions need to be met. Firstly, the timescale over which the asymmetry persists must be non-negligible compared to the timescale for giant collisions to occur. 
To estimate the catastrophic collision rate between Mars--sized bodies, we assume that there exists $N$ bodies of radius 3000~km and density 2700~kg/m$^3$ within a disk of radius 70~au and width 30~au \citep{Telesco2005} and that the bodies in the disk have a root-mean-square inclination of 0.156 \citep{Matra2019}. 
The collisional timescale of one body is given by 
\begin{equation}
    t_c = \frac{T}{4 \pi \tau},
\end{equation}
\noindent where $T$ is the orbital period and $\tau$ is the geometric optical depth of the $N$ bodies in the disk (i.e., their cross-sectional area divided by the cross-section of the annulus within which they reside) magnified by the gravitational focussing factor for collisions between equal sized bodies \citep{Wyatt1999}. Given that there are $N$ bodies in the disk, the timescale for a collision to occur is $t_c / N$. 

The collision of a Mars--sized body may leave a detectable asymmetry for up to 0.5~Myr \citep{Jackson2014}. In order for the probability of witnessing such an asymmetry to be above 1\%, the total number of Mars--sized bodies required in the disk is $>$500, corresponding to a total mass of $\sim$30~$M_\text{Earth}$.  

Secondly, in order for a giant collision to have likely occurred in the first place, the collisional timescale of the $N$ bodies, $t_c / N$, needs to be less than the age of the system. This requires the total number of Mars--sized bodies to be $>$800. Assuming that Mars--sized bodies represent the largest bodies in the disk, the limits from both conditions correspond to a total mass that is 30 times smaller than a disk with the surface density equal to the Minimum Mass Solar Nebula at the same location. A disk that supports this level of collision may therefore be plausible.

Although the data presented in this study alone may indeed be more consistent with the collision model, this model is not without observational challenges. \citet{Matra2017} found that the CO emission observed in the J = 2--1 transition is extended. As CO is expected to rapidly couple to the atomic gas disk, its broad extent suggests formation in a wide range of radii, which is inconsistent with the giant collision model in which the production of CO would occur through subsequent collisions at the giant collision site. 

Furthermore, it is possible that the event causing the dust clump is required to be more recent than the dust asymmetry timescale of 0.5~Myr. For example, \citet{Cataldi2018} estimated that such an event should have happened within the past 5,000~yr by deriving the total C\textsc{i} mass and assuming that all C\textsc{i} originates from CO at a constant rate and is not removed from the system. However, it should be noted that there is significant uncertainty associated with both the C\textsc{i} mass estimate and photodissociation rate of CO. It is also possible that not all collisions between Mars-sized bodies are able to produce a significant amount of observable debris, and so the number of Mars-sized bodies required for such giant impacts to be probable may be higher than estimated here.

\subsubsection{Tidal disruption}
Alternative to a giant impact, one scenario that could result in a broader radial distribution of material is the destruction of a Moon- or Mars-sized body by tidal disruption during a close encounter with a Neptune-sized planet \citep{Cataldi2018}. Compared to the giant impact scenario between Mars-sized bodies, tidal disruption by a Neptune-sized planet only requires close encounters to occur, which is made more likely by the significantly larger effect of gravitational focussing by such Neptune-sized planets, and loosens any requirement for large impact velocities which may be required to eject a large fraction of the mass as debris. \citet{Cataldi2018} estimated that it may be statistically plausible to witness such an event if the system includes a few Neptune-sized planets and several thousand Moon- to Mars-sized bodies. They argued that as the orbits of the post-impact material continue to gravitationally interact with the Neptune-sized planet within its Hill sphere, a radially broader clump of dust and gas may result from a tidal disruption event compared to a giant impact, however it may still not be radially broad enough to reproduce the observed CO clump \citep{Matra2019}.

\subsubsection{Secular perturbation}
A stationary dust clump may also originate from secular effects. \citet{Nesvold2015} were able to explain the central hole and warp of the disk with secular perturbations due to Beta~Pic~b, and found that a combination of a spiral density wave and vertical displacement wave originating from the eccentricity and inclination precessing at different rates at different semi-major axes could cause an asymmetric distribution in CO. Such an effect of secular perturbation could further increase the likelihood of witnessing collisions of large bodies by concentrating bodies in the disk, leading to a higher collision rate than assuming the bodies to be uniformly distributed throughout the disk. It may be worth further investigating how a clump of dust could form due to these forms of secular perturbation and how the observed amount of CO could be achieved.

\subsubsection{Eccentric disk}
It is also possible that the apparent stationary dust clump arises from the disk being eccentric. \citet{Cataldi2018} found that an initially eccentric disk would have circularised over the age of Beta Pic, and so any global eccentricity is likely due to secular perturbation by an unseen planet. However, such a planet is required to be highly eccentric \citep{Cataldi2018} and a high eccentricity dispersion in the disk is needed, which may conflict with the relatively narrow distribution of disk emission. 


\subsection{Future constraints}
\subsubsection{Disk observations}
Future observations may further tighten constraints on the motion of the dust clump and may allow us to confidently rule out models in which the dust clump is moving if they converge towards current estimates. 

\begin{figure}
    \centering
    \includegraphics[width=9cm]{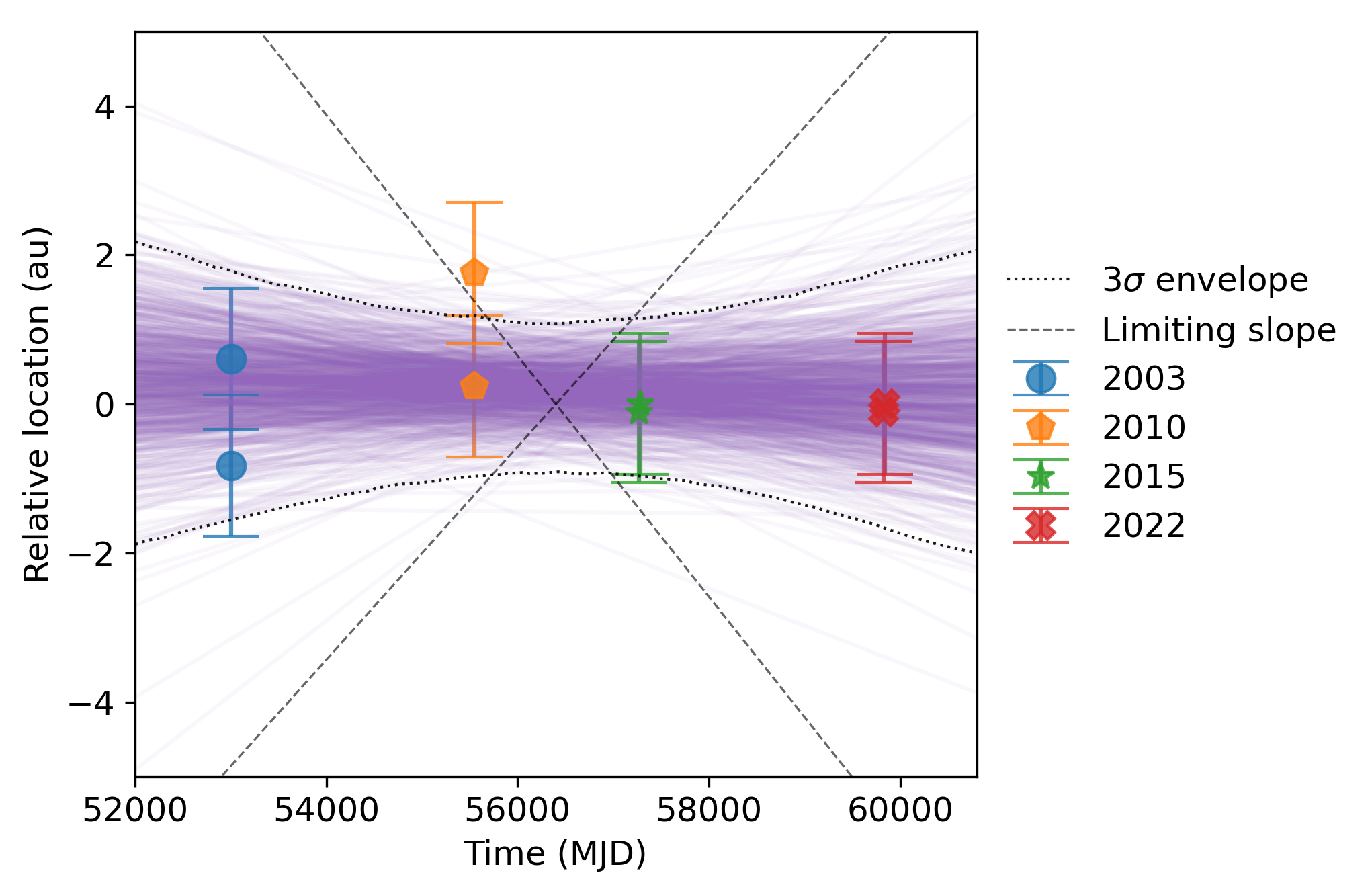}
    \caption{Same as Fig.~\ref{fig:line} but for a hypothetical scenario which involves two additional sets of observations in 2022 identical to those obtained in 2015. The dust clump locations are relative to 2022 Obs.~2. The limiting slopes correspond to the maximum and mimimum projected speed compatible with the resonance model. }
    \label{fig:line2}
\end{figure}

To estimate the constraints that could potentially be obtained with a new epoch of observation that is identical in sensitivity and resolution to the 2015 epoch, we duplicated the 2015 VISIR dataset, treating it as a new set of observations acquired on the same date in 2022, and repeated the analysis described in Section~\ref{sec:motion}. The results of a linear fit to the measured dust clump locations in this hypothetical scenario are shown in Figure~\ref{fig:line2} and Table~\ref{line2}.

\begin{table*}
    \centering
    \caption{Same as Table.~\ref{line} but for a hypothetical scenario which involves two additional sets of observations in 2022 identical to those obtained in 2015.}
    \label{line2}
    \begin{tabular}{l|c|c|c|c}
        \hline \hline
        & Median & 1$\sigma$ & 2$\sigma$ & 3$\sigma$ \\\hline
        Linear displacement speed (au/yr) & -0.011 & (-0.061, 0.040) & (-0.13, 0.11) & (-0.26, 0.24) \\
        Location at MJD = 0 (au) & -1.9 & (-6.0, 9.7) & (-16, 20) & (-37, 40) \\
        1$\sigma$ uncertainties on clump location (au) & 0.95 & (0.70, 1.4) & (0.55, 2.2) & (0.44, 4.0) \\
        \hline
    \end{tabular}
\end{table*}

These results imply that if there is no real displacement of the dust clump, a new observation identical in sensitivity and resolution could potentially constrain any motion of the dust clump to within 5~au over the 19~yr time baseline at the 3~sigma level, which is a factor of 3 tighter than current constraints on the proper motion. A resonant clump of planetesimals that is orbiting sufficiently slowly to fall into this range would require a perturbing planet outside 100~au, effectively ruling out the resonance scenario, leaving the giant collision model or other scenarios in which the clump is stationary as the more likely scenarios. 

We therefore expect to be able to set conclusive constraints with new observations. With the successful launch of JWST, a Cycle 1 Guaranteed Time Observations (GTO) program (proposal 1241, \citealp{Ressler2017jwst}) is scheduled to observe Beta~Pic with the F1140C (and F1065C) filter, which provide the most similar wavelength coverage to the observations used in this study. We expect such observations to further tighten current constraints on the motion and structure of the dust clump.

\subsubsection{Planet detection}
To gather additional lines of observational evidence on the nature of the dust clump, future observations may also wish to directly search for the presence of any planets that could be responsible for resonant trapping under the resonance model. 

JWST is expected to be able to detect a 2~$M_\text{Jup}$--planet at 50~au at the 80\% detection threshold with either the F444W filter on NIRCam or the F1550C filter on MIRI \citep{Carter2021} for an average system in the Beta Pic Moving Group. Given the relative proximity of Beta~Pic to the Earth, the detection probability may be higher for Beta~Pic, but the presence of disk emission may affect planet detection. The issue is less pronounced at 4~$\mu$m, where the disk flux is 2~mJy/arcsec$^2$ at 50~au along the NE arm \citep{Milli2014}. This is comparable to the predicted Jupiter--mass planet flux of $\sim$0.01~mJy at the age of Beta~Pic assuming cloud-free solar metallicity atmospheres \citep{Linder2019} over a PSF FWHM of 0.14$^{\prime\prime}$\footnote{\href{https://jwst-docs.stsci.edu/jwst-near-infrared-camera/nircam-predicted-performance/nircam-point-spread-functions}{https://jwst-docs.stsci.edu/jwst-near-infrared-camera/nircam-predicted-performance/nircam-point-spread-functions}}. At 15~$\mu$m, however, the disk flux is approximately 100~mJy/arcsec$^2$ at 50~au along the NE arm \citep{Telesco2005}, which is significantly brighter than the expected flux of a Jupiter--mass planet of $\sim$0.05~mJy \citep{Linder2019} over a PSF FWHM of 0.5$^{\prime\prime}$\footnote{\href{https://jwst-docs.stsci.edu/jwst-mid-infrared-instrument/miri-predicted-performance/miri-point-spread-functions}{https://jwst-docs.stsci.edu/jwst-mid-infrared-instrument/miri-predicted-performance/miri-point-spread-functions}}. The mass threshold for detection at this wavelength should be significantly higher than that in the absence of a disk. 

The edge-on perspective of Beta Pic implies that the projected separation of a planet is likely smaller than its orbital radius. For planetesimals with an eccentricity of 0.2 in the 2:1(l) resonance with a planet, the planet leads the clump of planetesimals by 85$^\circ$ \citep{Wyatt2003}. Given the range of orbital radii of the planet under the resonance model, the planet is expected to be seen at a projected distance of between 50 and 90~au. Detection of any planet within this range by JWST would be consistent with the perturbing planet required under the resonance model. Non--detection with the F444W filter would rule out the presence of a several $M_\text{Jup}$ perturber. If subsequent follow--up proper motion studies favour the resonance model, a non--detection would imply that a smaller perturbing planet with a slower maximum migration speed is required (Fig.~\ref{fig:migration}), which has more likely migrated by scattering planetesimals. 
A JWST GTO proposal is scheduled to observe the system with these and other filters (proposal 1411, \citealp{Stark2017jwst}) and may shed light on the presence or absence of any perturbing planets.

\subsection{Is the clump in Beta Pic unique?}
The Beta Pic debris disk is the only debris disk known to host such a significant clump of dust and gas. This invites the question of why clumps have not been observed more commonly in other debris disks. 

The uniqueness of the clump shall be interpreted in the context of its origin. If the dust clump originated from a giant impact, it may be that collisions that produce an observable amount of dust and gas that last long enough to be observed are indeed rare. Giant collisions may still be common, but their signatures such as a dust clump may be transient. The probability of witnessing a giant impact may have been enhanced in Beta Pic due to its young age and large mass, allowing for a higher number of larger bodies to exist in the disk to undergo such collisions. 

If the dust clump originated from resonantly trapped planetesimals, it is possible that the probability of witnessing resonant trapping is enhanced in Beta Pic once again due to its large mass, which provides a sufficient amount of primordial gas or planetesimals to be scattered for massive planets to migrate in the first place. Although planets are believed to be prevalent, the conditions required to sustain outwards migration may be relatively stringent. For example, the probability of trapping planetesimals may be lower if the planetesimals were initially on eccentric orbits \citep{Reche2008}. Even if dust or gas clumps due to resonant trapping are more common than has been observed, they may be more difficult to witness in other debris disks that are not as young, bright and nearby as Beta Pic. 

As more high resolution images become available, discovery of dust or gas clumps in other debris disks or the lack of it may offer insight into the uniqueness of this feature in the context of its origin.

\section{Conclusions}
\label{sec:conclusions}
We presented VLT/VISIR observations of Beta~Pic in the mid-IR and compared these observations to earlier epochs to constrain the motion of the dust clump in the SW arm of the debris disk. We found that over the 12-yr period spanned by the observations, the projected displacement along the disk's major axis is $0.2^{+1.3}_{-1.4}$~au away from the star based on the median and $1\sigma$ range of the posterior distribution, and between -10~au and +11~au away from the star at the $3\sigma$ level.


A number of models have been proposed to explain the origin of the dust clump. Our results appear to disfavour models in which the dust clump is moving, including a resonant planetesimal model \citep{Wyatt2003} in which the dust clump orbits the star, but does not conclusively rule them out. 

However, if the origin of the dust clump is dust generated from the collision of planetesimals trapped in resonance with an undetected perturbing planet, as proposed under the resonance model, we rule out the presence of such a perturbing planet within 70~au based on the $3\sigma$ limits of the clump's motion. 
We also estimate based on the age of Beta~Pic that the migration speed of the perturbing planet required to excite the planetesimals from a circular orbit to this eccentricity is at least 0.3~au/Myr, and that the minimum mass of the perturbing planet required to capture the planetesimals into a 2:1 resonance under this migration speed is approximately 10~$M_\text{Earth}$. 
Such a migration scenario could be achieved by a planet with a mass of 3~M$_\text{Jupiter}$ or greater migrating under either planetesimal--driven migration or gas--driven migration, or by a smaller Uranus--mass planet via planetesimal--driven migration. JWST may be able to detect planets that could be responsible for resonant trapping under the resonance model. 


Alternative scenarios in which the dust clump is stationary, such as from the collision \citep{Jackson2014} or tidal disruption \citep{Cataldi2018} of a planet--sized body, or from a region of enhanced density and collision from secular perturbation \citep{Nesvold2015}, may be more consistent with the results in this study alone. As the giant impact and tidal disruption models have been challenged by CO observations \citep{Matra2017}, future monitoring of the clump motion both in the mid-infrared and in CO may provide more evidence pointing towards the more likely scenario. Any model would also be required to explain the C\textsc{i} observations \citep{Cataldi2018}. We expect that the incorporation of new JWST observations into this analysis may be able to provide significantly tighter constraints on the dust clump's motion, potentially providing conclusive evidence against the resonant model. 

Future studies may also wish to further investigate the formation of a dust clump via secular perturbation \citep{Nesvold2015}
and the expected clump morphology and motion under this model. Any detection of planets that satisfy the conditions required by the resonance model may also shed light on the origin of the dust clump.

\section*{Acknowledgements}
YH acknowledges funding from the Gates Cambridge Trust. This research made use of NASA's Astrophysics Data System; the \textsc{IPython} package \citep{ipython}; \textsc{SciPy} \citep{scipy}; \textsc{NumPy} \citep{numpy}; \textsc{matplotlib} \citep{matplotlib}; and \textsc{Astropy}, a community-developed core Python package for Astronomy \citep{astropy}.

\section*{Data Availability}
The T-ReCS data used in this study are available on the Gemini Observatory Archive under programme IDs GS-2003B-Q-14 and GS-2010B-Q-50. The VISIR data used in this study are available on the ESO Archive under programme ID 095.C-0425(A).

\bibliographystyle{mnras}
\bibliography{references}

\begin{thebibliography}{}
\makeatletter
\relax
\def\mn@urlcharsother{\let\do\@makeother \do\$\do\&\do\#\do\^\do\_\do\%\do\~}
\def\mn@doi{\begingroup\mn@urlcharsother \@ifnextchar [ {\mn@doi@}
  {\mn@doi@[]}}
\def\mn@doi@[#1]#2{\def\@tempa{#1}\ifx\@tempa\@empty \href
  {http://dx.doi.org/#2} {doi:#2}\else \href {http://dx.doi.org/#2} {#1}\fi
  \endgroup}
\def\mn@eprint#1#2{\mn@eprint@#1:#2::\@nil}
\def\mn@eprint@arXiv#1{\href {http://arxiv.org/abs/#1} {{\tt arXiv:#1}}}
\def\mn@eprint@dblp#1{\href {http://dblp.uni-trier.de/rec/bibtex/#1.xml}
  {dblp:#1}}
\def\mn@eprint@#1:#2:#3:#4\@nil{\def\@tempa {#1}\def\@tempb {#2}\def\@tempc
  {#3}\ifx \@tempc \@empty \let \@tempc \@tempb \let \@tempb \@tempa \fi \ifx
  \@tempb \@empty \def\@tempb {arXiv}\fi \@ifundefined
  {mn@eprint@\@tempb}{\@tempb:\@tempc}{\expandafter \expandafter \csname
  mn@eprint@\@tempb\endcsname \expandafter{\@tempc}}}

\bibitem[\protect\citeauthoryear{{Apai}, {Schneider}, {Grady}, {Wyatt},
  {Lagrange}, {Kuchner}, {Stark}  \& {Lubow}}{{Apai} et~al.}{2015}]{Apai2015}
{Apai} D.,  {Schneider} G.,  {Grady} C.~A.,  {Wyatt} M.~C.,  {Lagrange} A.-M.,
  {Kuchner} M.~J.,  {Stark} C.~J.,   {Lubow} S.~H.,  2015, \mn@doi [\apj]
  {10.1088/0004-637X/800/2/136}, \href
  {https://ui.adsabs.harvard.edu/abs/2015ApJ...800..136A} {800, 136}

\bibitem[\protect\citeauthoryear{{Astropy Collaboration} et~al.,}{{Astropy
  Collaboration} et~al.}{2013}]{astropy}
{Astropy Collaboration} et~al., 2013, \mn@doi [\aap]
  {10.1051/0004-6361/201322068}, \href
  {http://adsabs.harvard.edu/abs/2013A%26A...558A..33A} {558, A33}

\bibitem[\protect\citeauthoryear{{Bate}, {Lubow}, {Ogilvie}  \&
  {Miller}}{{Bate} et~al.}{2003}]{Bate2003}
{Bate} M.~R.,  {Lubow} S.~H.,  {Ogilvie} G.~I.,   {Miller} K.~A.,  2003,
  \mn@doi [\mnras] {10.1046/j.1365-8711.2003.06406.x}, \href
  {https://ui.adsabs.harvard.edu/abs/2003MNRAS.341..213B} {341, 213}

\bibitem[\protect\citeauthoryear{{Binks} \& {Jeffries}}{{Binks} \&
  {Jeffries}}{2014}]{Binks2014}
{Binks} A.~S.,  {Jeffries} R.~D.,  2014, \mn@doi [\mnras]
  {10.1093/mnrasl/slt141}, \href
  {https://ui.adsabs.harvard.edu/abs/2014MNRAS.438L..11B} {438, L11}

\bibitem[\protect\citeauthoryear{{Bitsch} \& {Johansen}}{{Bitsch} \&
  {Johansen}}{2016}]{Bitsch2016}
{Bitsch} B.,  {Johansen} A.,  2016, \mn@doi [\aap]
  {10.1051/0004-6361/201527676}, \href
  {https://ui.adsabs.harvard.edu/abs/2016A&A...590A.101B} {590, A101}

\bibitem[\protect\citeauthoryear{{Bitsch} \& {Kley}}{{Bitsch} \&
  {Kley}}{2011}]{Bitsch2011}
{Bitsch} B.,  {Kley} W.,  2011, \mn@doi [\aap] {10.1051/0004-6361/201117202},
  \href {https://ui.adsabs.harvard.edu/abs/2011A&A...536A..77B} {536, A77}

\bibitem[\protect\citeauthoryear{{Bonsor}, {Raymond}, {Augereau}  \&
  {Ormel}}{{Bonsor} et~al.}{2014}]{Bonsor2014}
{Bonsor} A.,  {Raymond} S.~N.,  {Augereau} J.-C.,   {Ormel} C.~W.,  2014,
  \mn@doi [\mnras] {10.1093/mnras/stu721}, \href
  {https://ui.adsabs.harvard.edu/abs/2014MNRAS.441.2380B} {441, 2380}

\bibitem[\protect\citeauthoryear{{Carter} et~al.,}{{Carter}
  et~al.}{2021}]{Carter2021}
{Carter} A.~L.,  et~al., 2021, \mn@doi [\mnras] {10.1093/mnras/staa3579}, \href
  {https://ui.adsabs.harvard.edu/abs/2021MNRAS.501.1999C} {501, 1999}

\bibitem[\protect\citeauthoryear{{Cataldi} et~al.,}{{Cataldi}
  et~al.}{2018}]{Cataldi2018}
{Cataldi} G.,  et~al., 2018, \mn@doi [\apj] {10.3847/1538-4357/aac5f3}, \href
  {https://ui.adsabs.harvard.edu/abs/2018ApJ...861...72C} {861, 72}

\bibitem[\protect\citeauthoryear{{Chavez-Dagostino} et~al.,}{{Chavez-Dagostino}
  et~al.}{2016}]{Chavez2016}
{Chavez-Dagostino} M.,  et~al., 2016, \mn@doi [\mnras] {10.1093/mnras/stw1363},
  \href {https://ui.adsabs.harvard.edu/abs/2016MNRAS.462.2285C} {462, 2285}

\bibitem[\protect\citeauthoryear{{Crida}, {Morbidelli}  \& {Masset}}{{Crida}
  et~al.}{2006}]{Crida2006}
{Crida} A.,  {Morbidelli} A.,   {Masset} F.,  2006, \mn@doi [\icarus]
  {10.1016/j.icarus.2005.10.007}, \href
  {https://ui.adsabs.harvard.edu/abs/2006Icar..181..587C} {181, 587}

\bibitem[\protect\citeauthoryear{{Crida}, {Masset}  \& {Morbidelli}}{{Crida}
  et~al.}{2009}]{Crida2009}
{Crida} A.,  {Masset} F.,   {Morbidelli} A.,  2009, \mn@doi [\apjl]
  {10.1088/0004-637X/705/2/L148}, \href
  {https://ui.adsabs.harvard.edu/abs/2009ApJ...705L.148C} {705, L148}

\bibitem[\protect\citeauthoryear{{Crifo}, {Vidal-Madjar}, {Lallement}, {Ferlet}
   \& {Gerbaldi}}{{Crifo} et~al.}{1997}]{Crifo1997}
{Crifo} F.,  {Vidal-Madjar} A.,  {Lallement} R.,  {Ferlet} R.,   {Gerbaldi} M.,
   1997, \aap, \href {https://ui.adsabs.harvard.edu/abs/1997A&A...320L..29C}
  {320, L29}

\bibitem[\protect\citeauthoryear{{Dent} et~al.,}{{Dent}
  et~al.}{2014}]{Dent2014}
{Dent} W.~R.~F.,  et~al., 2014, \mn@doi [Science] {10.1126/science.1248726},
  \href {https://ui.adsabs.harvard.edu/abs/2014Sci...343.1490D} {343, 1490}

\bibitem[\protect\citeauthoryear{{Fernandez} \& {Ip}}{{Fernandez} \&
  {Ip}}{1984}]{Fernandez1984}
{Fernandez} J.~A.,  {Ip} W.~H.,  1984, \mn@doi [\icarus]
  {10.1016/0019-1035(84)90101-5}, \href
  {https://ui.adsabs.harvard.edu/abs/1984Icar...58..109F} {58, 109}

\bibitem[\protect\citeauthoryear{{Foreman-Mackey}, {Hogg}, {Lang}  \&
  {Goodman}}{{Foreman-Mackey} et~al.}{2013}]{emcee}
{Foreman-Mackey} D.,  {Hogg} D.~W.,  {Lang} D.,   {Goodman} J.,  2013, \mn@doi
  [\pasp] {10.1086/670067}, \href
  {https://ui.adsabs.harvard.edu/abs/2013PASP..125..306F} {125, 306}

\bibitem[\protect\citeauthoryear{{Gaia Collaboration} et~al.,}{{Gaia
  Collaboration} et~al.}{2018}]{Gaia2018}
{Gaia Collaboration} et~al., 2018, \mn@doi [\aap]
  {10.1051/0004-6361/201833051}, \href
  {https://ui.adsabs.harvard.edu/abs/2018A&A...616A...1G} {616, A1}

\bibitem[\protect\citeauthoryear{{Golimowski} et~al.,}{{Golimowski}
  et~al.}{2006}]{Golimowski2006}
{Golimowski} D.~A.,  et~al., 2006, \mn@doi [\aj] {10.1086/503801}, \href
  {https://ui.adsabs.harvard.edu/abs/2006AJ....131.3109G} {131, 3109}

\bibitem[\protect\citeauthoryear{{Grigorieva}, {Artymowicz}  \&
  {Th{\'e}bault}}{{Grigorieva} et~al.}{2007}]{Grigorieva2007}
{Grigorieva} A.,  {Artymowicz} P.,   {Th{\'e}bault} P.,  2007, \mn@doi [\aap]
  {10.1051/0004-6361:20065210}, \href
  {https://ui.adsabs.harvard.edu/abs/2007A&A...461..537G} {461, 537}

\bibitem[\protect\citeauthoryear{{Haworth}, {Cadman}, {Meru}, {Hall},
  {Albertini}, {Forgan}, {Rice}  \& {Owen}}{{Haworth}
  et~al.}{2020}]{Haworth2020}
{Haworth} T.~J.,  {Cadman} J.,  {Meru} F.,  {Hall} C.,  {Albertini} E.,
  {Forgan} D.,  {Rice} K.,   {Owen} J.~E.,  2020, \mn@doi [\mnras]
  {10.1093/mnras/staa883}, \href
  {https://ui.adsabs.harvard.edu/abs/2020MNRAS.494.4130H} {494, 4130}

\bibitem[\protect\citeauthoryear{{Hayashi}}{{Hayashi}}{1981}]{Hayashi1981}
{Hayashi} C.,  1981, \mn@doi [Progress of Theoretical Physics Supplement]
  {10.1143/PTPS.70.35}, \href
  {https://ui.adsabs.harvard.edu/abs/1981PThPS..70...35H} {70, 35}

\bibitem[\protect\citeauthoryear{{Houk}}{{Houk}}{1978}]{Houk1978}
{Houk} N.,  1978, {Michigan catalogue of two-dimensional spectral types for the
  HD stars}.
Stellar catalogue

\bibitem[\protect\citeauthoryear{{Hughes}, {Duch{\^e}ne}  \&
  {Matthews}}{{Hughes} et~al.}{2018}]{Hughes2018}
{Hughes} A.~M.,  {Duch{\^e}ne} G.,   {Matthews} B.~C.,  2018, \mn@doi [ARAA]
  {10.1146/annurev-astro-081817-052035}, \href
  {https://ui.adsabs.harvard.edu/abs/2018ARA&A..56..541H} {56, 541}

\bibitem[\protect\citeauthoryear{Hunter}{Hunter}{2007}]{matplotlib}
Hunter J.~D.,  2007, Computing In Science \& Engineering, 9, 90

\bibitem[\protect\citeauthoryear{{Ida}, {Bryden}, {Lin}  \& {Tanaka}}{{Ida}
  et~al.}{2000}]{Ida2000}
{Ida} S.,  {Bryden} G.,  {Lin} D.~N.~C.,   {Tanaka} H.,  2000, \mn@doi [\apj]
  {10.1086/308720}, \href
  {https://ui.adsabs.harvard.edu/abs/2000ApJ...534..428I} {534, 428}

\bibitem[\protect\citeauthoryear{{Jackson}, {Wyatt}, {Bonsor}  \&
  {Veras}}{{Jackson} et~al.}{2014}]{Jackson2014}
{Jackson} A.~P.,  {Wyatt} M.~C.,  {Bonsor} A.,   {Veras} D.,  2014, \mn@doi
  [\mnras] {10.1093/mnras/stu476}, \href
  {https://ui.adsabs.harvard.edu/abs/2014MNRAS.440.3757J} {440, 3757}

\bibitem[\protect\citeauthoryear{Jones, Oliphant, Peterson  \& Others}{Jones
  et~al.}{2001}]{scipy}
Jones E.,  Oliphant T.,  Peterson P.,   Others 2001, {SciPy}: Open source
  scientific tools for Python, \url {http://www.scipy.org/}

\bibitem[\protect\citeauthoryear{{Kirsh}, {Duncan}, {Brasser}  \&
  {Levison}}{{Kirsh} et~al.}{2009}]{Kirsh2009}
{Kirsh} D.~R.,  {Duncan} M.,  {Brasser} R.,   {Levison} H.~F.,  2009, \mn@doi
  [\icarus] {10.1016/j.icarus.2008.05.028}, \href
  {https://ui.adsabs.harvard.edu/abs/2009Icar..199..197K} {199, 197}

\bibitem[\protect\citeauthoryear{{Kley} \& {Nelson}}{{Kley} \&
  {Nelson}}{2012}]{Kley2012}
{Kley} W.,  {Nelson} R.~P.,  2012, \mn@doi [\araa]
  {10.1146/annurev-astro-081811-125523}, \href
  {https://ui.adsabs.harvard.edu/abs/2012ARA&A..50..211K} {50, 211}

\bibitem[\protect\citeauthoryear{{Krivov} \& {Wyatt}}{{Krivov} \&
  {Wyatt}}{2021}]{Krivov2021}
{Krivov} A.~V.,  {Wyatt} M.~C.,  2021, \mn@doi [\mnras]
  {10.1093/mnras/staa2385}, \href
  {https://ui.adsabs.harvard.edu/abs/2021MNRAS.500..718K} {500, 718}

\bibitem[\protect\citeauthoryear{{Lagrange} et~al.,}{{Lagrange}
  et~al.}{2009}]{Lagrange2009}
{Lagrange} A.~M.,  et~al., 2009, \mn@doi [\aap] {10.1051/0004-6361:200811325},
  \href {https://ui.adsabs.harvard.edu/abs/2009A&A...493L..21L} {493, L21}

\bibitem[\protect\citeauthoryear{{Larwood} \& {Kalas}}{{Larwood} \&
  {Kalas}}{2001}]{Larwood2001}
{Larwood} J.~D.,  {Kalas} P.~G.,  2001, \mn@doi [\mnras]
  {10.1046/j.1365-8711.2001.04212.x}, \href
  {https://ui.adsabs.harvard.edu/abs/2001MNRAS.323..402L} {323, 402}

\bibitem[\protect\citeauthoryear{{Li}, {Telesco}  \& {Wright}}{{Li}
  et~al.}{2012}]{Li2012}
{Li} D.,  {Telesco} C.~M.,   {Wright} C.~M.,  2012, \mn@doi [\apj]
  {10.1088/0004-637X/759/2/81}, \href
  {https://ui.adsabs.harvard.edu/abs/2012ApJ...759...81L} {759, 81}

\bibitem[\protect\citeauthoryear{{Lin} \& {Papaloizou}}{{Lin} \&
  {Papaloizou}}{1986}]{Lin1986}
{Lin} D.~N.~C.,  {Papaloizou} J.,  1986, \mn@doi [\apj] {10.1086/164426}, \href
  {https://ui.adsabs.harvard.edu/abs/1986ApJ...307..395L} {307, 395}

\bibitem[\protect\citeauthoryear{{Linder}, {Mordasini}, {Molli{\`e}re},
  {Marleau}, {Malik}, {Quanz}  \& {Meyer}}{{Linder} et~al.}{2019}]{Linder2019}
{Linder} E.~F.,  {Mordasini} C.,  {Molli{\`e}re} P.,  {Marleau} G.-D.,  {Malik}
  M.,  {Quanz} S.~P.,   {Meyer} M.~R.,  2019, \mn@doi [\aap]
  {10.1051/0004-6361/201833873}, \href
  {https://ui.adsabs.harvard.edu/abs/2019A&A...623A..85L} {623, A85}

\bibitem[\protect\citeauthoryear{{MacGregor} et~al.,}{{MacGregor}
  et~al.}{2017}]{MacGregor2017Fom}
{MacGregor} M.~A.,  et~al., 2017, \mn@doi [\apj] {10.3847/1538-4357/aa71ae},
  \href {https://ui.adsabs.harvard.edu/abs/2017ApJ...842....8M} {842, 8}

\bibitem[\protect\citeauthoryear{{Martin}, {Lubow}, {Pringle}  \&
  {Wyatt}}{{Martin} et~al.}{2007}]{Martin2007}
{Martin} R.~G.,  {Lubow} S.~H.,  {Pringle} J.~E.,   {Wyatt} M.~C.,  2007,
  \mn@doi [\mnras] {10.1111/j.1365-2966.2007.11906.x}, \href
  {https://ui.adsabs.harvard.edu/abs/2007MNRAS.378.1589M} {378, 1589}

\bibitem[\protect\citeauthoryear{{Matr{\`a}} et~al.,}{{Matr{\`a}}
  et~al.}{2017}]{Matra2017}
{Matr{\`a}} L.,  et~al., 2017, \mn@doi [\mnras] {10.1093/mnras/stw2415}, \href
  {https://ui.adsabs.harvard.edu/abs/2017MNRAS.464.1415M} {464, 1415}

\bibitem[\protect\citeauthoryear{{Matr{\`a}}, {Wyatt}, {Wilner}, {Dent},
  {Marino}, {Kennedy}  \& {Milli}}{{Matr{\`a}} et~al.}{2019}]{Matra2019}
{Matr{\`a}} L.,  {Wyatt} M.~C.,  {Wilner} D.~J.,  {Dent} W.~R.~F.,  {Marino}
  S.,  {Kennedy} G.~M.,   {Milli} J.,  2019, \mn@doi [\aj]
  {10.3847/1538-3881/ab06c0}, \href
  {https://ui.adsabs.harvard.edu/abs/2019AJ....157..135M} {157, 135}

\bibitem[\protect\citeauthoryear{{Matr{\`a}} et~al.,}{{Matr{\`a}}
  et~al.}{2020}]{Matra2020}
{Matr{\`a}} L.,  et~al., 2020, \mn@doi [\apj] {10.3847/1538-4357/aba0a4}, \href
  {https://ui.adsabs.harvard.edu/abs/2020ApJ...898..146M} {898, 146}

\bibitem[\protect\citeauthoryear{{Milli} et~al.,}{{Milli}
  et~al.}{2014}]{Milli2014}
{Milli} J.,  et~al., 2014, \mn@doi [\aap] {10.1051/0004-6361/201323130}, \href
  {https://ui.adsabs.harvard.edu/abs/2014A&A...566A..91M} {566, A91}

\bibitem[\protect\citeauthoryear{{Nesvold} \& {Kuchner}}{{Nesvold} \&
  {Kuchner}}{2015}]{Nesvold2015}
{Nesvold} E.~R.,  {Kuchner} M.~J.,  2015, \mn@doi [\apj]
  {10.1088/0004-637X/815/1/61}, \href
  {https://ui.adsabs.harvard.edu/abs/2015ApJ...815...61N} {815, 61}

\bibitem[\protect\citeauthoryear{{Ozernoy}, {Gorkavyi}, {Mather}  \&
  {Taidakova}}{{Ozernoy} et~al.}{2000}]{Ozernoy2000}
{Ozernoy} L.~M.,  {Gorkavyi} N.~N.,  {Mather} J.~C.,   {Taidakova} T.~A.,
  2000, \mn@doi [\apjl] {10.1086/312779}, \href
  {https://ui.adsabs.harvard.edu/abs/2000ApJ...537L.147O} {537, L147}

\bibitem[\protect\citeauthoryear{{Paardekooper}, {Dong}, {Duffell}, {Fung},
  {Masset}, {Ogilvie}  \& {Tanaka}}{{Paardekooper}
  et~al.}{2022}]{Paardekooper2022}
{Paardekooper} S.-J.,  {Dong} R.,  {Duffell} P.,  {Fung} J.,  {Masset} F.~S.,
  {Ogilvie} G.,   {Tanaka} H.,  2022, arXiv e-prints, \href
  {https://ui.adsabs.harvard.edu/abs/2022arXiv220309595P} {p. arXiv:2203.09595}

\bibitem[\protect\citeauthoryear{{Pantin}, {Lagage}  \& {Artymowicz}}{{Pantin}
  et~al.}{1997}]{Pantin1997}
{Pantin} E.,  {Lagage} P.~O.,   {Artymowicz} P.,  1997, \aap, \href
  {https://ui.adsabs.harvard.edu/abs/1997A&A...327.1123P} {327, 1123}

\bibitem[\protect\citeauthoryear{P\'erez \& Granger}{P\'erez \&
  Granger}{2007}]{ipython}
P\'erez F.,  Granger B.~E.,  2007, \mn@doi [Computing in Science and
  Engineering] {10.1109/MCSE.2007.53}, 9, 21

\bibitem[\protect\citeauthoryear{{Poulton}, {Greaves}  \& {Collier
  Cameron}}{{Poulton} et~al.}{2006}]{Poulton2006}
{Poulton} C.~J.,  {Greaves} J.~S.,   {Collier Cameron} A.,  2006, \mn@doi
  [\mnras] {10.1111/j.1365-2966.2006.10708.x}, \href
  {https://ui.adsabs.harvard.edu/abs/2006MNRAS.372...53P} {372, 53}

\bibitem[\protect\citeauthoryear{{Quillen} \& {Thorndike}}{{Quillen} \&
  {Thorndike}}{2002}]{Quillen2002}
{Quillen} A.~C.,  {Thorndike} S.,  2002, \mn@doi [\apjl] {10.1086/344708},
  \href {https://ui.adsabs.harvard.edu/abs/2002ApJ...578L.149Q} {578, L149}

\bibitem[\protect\citeauthoryear{{Reche}, {Beust}, {Augereau}  \&
  {Absil}}{{Reche} et~al.}{2008}]{Reche2008}
{Reche} R.,  {Beust} H.,  {Augereau} J.~C.,   {Absil} O.,  2008, \mn@doi [\aap]
  {10.1051/0004-6361:20077934}, \href
  {https://ui.adsabs.harvard.edu/abs/2008A&A...480..551R} {480, 551}

\bibitem[\protect\citeauthoryear{{Ressler}, {Choquet}  \& {Serabyn}}{{Ressler}
  et~al.}{2017}]{Ressler2017jwst}
{Ressler} M.~E.,  {Choquet} E.,   {Serabyn} G.,  2017, {MIRI Coronagraphic
  Imaging of exoplanets}, JWST Proposal. Cycle 1

\bibitem[\protect\citeauthoryear{{Stark}, {Clampin}, {Mountain}, {Perrin},
  {Pueyo}, {Rajan}  \& {Soummer}}{{Stark} et~al.}{2017}]{Stark2017jwst}
{Stark} C.~C.,  {Clampin} M.,  {Mountain} M.,  {Perrin} M.,  {Pueyo} L.,
  {Rajan} A.,   {Soummer} R.,  2017, {Coronagraphy of the Debris Disk Archetype
  Beta Pictoris}, JWST Proposal. Cycle 1

\bibitem[\protect\citeauthoryear{{Tanaka}, {Takeuchi}  \& {Ward}}{{Tanaka}
  et~al.}{2002}]{Tanaka2002}
{Tanaka} H.,  {Takeuchi} T.,   {Ward} W.~R.,  2002, \mn@doi [\apj]
  {10.1086/324713}, \href
  {https://ui.adsabs.harvard.edu/abs/2002ApJ...565.1257T} {565, 1257}

\bibitem[\protect\citeauthoryear{{Telesco} et~al.,}{{Telesco}
  et~al.}{2005}]{Telesco2005}
{Telesco} C.~M.,  et~al., 2005, \mn@doi [\nat] {10.1038/nature03255}, \href
  {https://ui.adsabs.harvard.edu/abs/2005Natur.433..133T} {433, 133}

\bibitem[\protect\citeauthoryear{{Thebault} \& {Kral}}{{Thebault} \&
  {Kral}}{2018}]{Thebault2018}
{Thebault} P.,  {Kral} Q.,  2018, \mn@doi [\aap] {10.1051/0004-6361/201731819},
  \href {https://ui.adsabs.harvard.edu/abs/2018A&A...609A..98T} {609, A98}

\bibitem[\protect\citeauthoryear{Van Der~Walt, Colbert  \& Varoquaux}{Van
  Der~Walt et~al.}{2011}]{numpy}
Van Der~Walt S.,  Colbert S.~C.,   Varoquaux G.,  2011, Computing in Science \&
  Engineering, 13, 22

\bibitem[\protect\citeauthoryear{{Ward}}{{Ward}}{1997}]{Ward1997}
{Ward} W.~R.,  1997, \mn@doi [\icarus] {10.1006/icar.1996.5647}, \href
  {https://ui.adsabs.harvard.edu/abs/1997Icar..126..261W} {126, 261}

\bibitem[\protect\citeauthoryear{{Wilner}, {Holman}, {Kuchner}  \&
  {Ho}}{{Wilner} et~al.}{2002}]{Wilner2002}
{Wilner} D.~J.,  {Holman} M.~J.,  {Kuchner} M.~J.,   {Ho} P.~T.~P.,  2002,
  \mn@doi [\apjl] {10.1086/340691}, \href
  {https://ui.adsabs.harvard.edu/abs/2002ApJ...569L.115W} {569, L115}

\bibitem[\protect\citeauthoryear{{Wyatt}}{{Wyatt}}{2003}]{Wyatt2003}
{Wyatt} M.~C.,  2003, \mn@doi [\apj] {10.1086/379064}, \href
  {https://ui.adsabs.harvard.edu/abs/2003ApJ...598.1321W} {598, 1321}

\bibitem[\protect\citeauthoryear{{Wyatt}, {Dermott}, {Telesco}, {Fisher},
  {Grogan}, {Holmes}  \& {Pi{\~n}a}}{{Wyatt} et~al.}{1999}]{Wyatt1999}
{Wyatt} M.~C.,  {Dermott} S.~F.,  {Telesco} C.~M.,  {Fisher} R.~S.,  {Grogan}
  K.,  {Holmes} E.~K.,   {Pi{\~n}a} R.~K.,  1999, \mn@doi [\apj]
  {10.1086/308093}, \href
  {https://ui.adsabs.harvard.edu/abs/1999ApJ...527..918W} {527, 918}

\bibitem[\protect\citeauthoryear{de Wit \& the VISIR~IOT}{de~Wit \& the
  VISIR~IOT}{2020}]{deWit2020}
de Wit W.,  the VISIR~IOT 2020, Very Large Telescope Paranal Science Operations
  VISIR User Manual.
\url
  {https://www.eso.org/sci/facilities/paranal/instruments/visir/doc/VLT-MAN-ESO-14300-3514_2020-03-03.pdf}

\makeatother
\end{thebibliography}

\appendix

\section{Subtract first or convolve first}
\label{appendix:subtraction}
Here, we illustrate that when comparing the dust clump using two different images, convolving the two original images with each other's PSF and then rotationally subtracting the image to isolate the clump image is the preferred procedure when compared to first isolating the clump images and then convolving with each other's PSF. This is fundamentally because the former procedure reproduces the effects of any PSF asymmetries in both resultant clump images, allowing for a fair comparison between the two. 

Consider an underlying source distribution, $D(\vb* x)$, imaged independently with PSFs $P_1(\vb* x)$ and $P_2(\vb* x)$ to give images $I_1(\vb* x)$ and $I_2(\vb* x)$:
\begin{equation}
    I_1(\vb* x) = D(\vb* x) * P_1(\vb* x),
\end{equation}
\begin{equation}
    I_2(\vb* x) = D(\vb* x) * P_2(\vb* x),
\end{equation}

\noindent where the star symbol denotes convolution. When convolving before subtracting (i.e., the former case), both images give
\begin{equation}
    I_{\text{conv, subt}} = D(\vb* x) * P_1(\vb* x) * P_2(\vb* x) - D(-\vb* x) * P_1(-\vb* x) * P_2(-\vb* x).
\end{equation}
\noindent Even though the two observations, $I_1(\vb* x)$ and $I_2(\vb* x)$, are different, the procedure in theory gives the same outcome after applying this procedure, which is desired since they are observations of the same underlying source distribution. 

When subtracting before convolving (i.e., the latter case), the two images give different outcomes, which are given by
\begin{equation}
    I_{1, \text{subt, conv}} = D(\vb* x) * P_1(\vb* x) * P_2(\vb* x) - D(-\vb* x) * P_1(-\vb* x) * P_2(\vb* x),
\end{equation}
\begin{equation}
    I_{2, \text{subt, conv}} = D(\vb* x) * P_1(\vb* x) * P_2(\vb* x) - D(-\vb* x) * P_1(\vb* x) * P_2(-\vb* x).
\end{equation}

We see that unless the two PSFs are rotationally symmetric, i.e., $P(\vb* x) = P(-\vb* x)$, the latter procedure gives different outputs following the same procedure even when imaging the same underlying source distribution, therefore not allowing for a fair comparison between two underlying distributions. For this reason, we choose to apply the former procedure to account for both PSFs when isolating the dust clump for comparison.

\section{Planet migration scenarios}
\label{appendix:migration}
Two major categories of migration mechanisms may be responsible for the outwards migration of a planet, including planetesimal--driven migration and gas--driven migration. Most models of planet migration assume that planets migrate inwards, but outward migration is also possible for both planetesimal--driven \citep{Fernandez1984, Bonsor2014} and gas--driven migration \citep{Martin2007, Crida2009, Bitsch2011} under certain conditions \citep{Bitsch2016, Paardekooper2022}. 
Here we use the predictions for inward migration as representative of the rates achieved for outward migration.

Under planetesimal--driven migration, a planet exchanges angular momentum by scattering planetesimals \citep{Kirsh2009}. 
\citet{Ida2000} derived a relationship for the migration rate, 
\begin{equation}
    \bigg | \frac{dr}{dt} \bigg | = 4 \pi \Sigma \frac{r^3}{M_* T},
\end{equation}
\noindent where $r$ is the orbital radius of the planet, $\Sigma$ is the surface density of the disk of planetesimals, M$_*$ is the mass of the star and T is the orbital period of the planet. 

For a planet starting at 62~au and migrating at above 0.3~au/Myr, a surface density of above $2.2 \times 10^{-5}$~M$_\text{Earth}/\text{au}^2$ is required at this radius. For a planetesimal disk of uniform surface density extending from 0 to 100~au, this would correspond to a total planetesimal mass of 0.7~M$_\text{Earth}$. This is well below the theoretical maximum possible planetesimal mass of 600~M$_\text{Earth}$ for Beta~Pic, which is estimated by assuming a maximum disk mass of 0.1 times the stellar mass in order to be gravitational stable \citep{Haworth2020} and a dust-to-gas ratio of 0.01 \citep{Krivov2021}. Planetesimal--driven migration could therefore plausibly deliver the migration rate required under the resonance model. 


Another potential mechanism at play is gas--driven migration, in which planets migrate through interactions with a gaseous disk \citep{Kley2012}. The requirement for high densities of gas implies that this is expected to occur primarily at relatively early phases of the planetary system's formation and evolution. At low planet masses, the disk density is unaffected by the planet and migration occurs in a linear regime know as Type I migration \citep{Ward1997}. The migration speed is given by
\begin{equation}
    \label{eq:typeI}
    v_\text{I} = \bigg | \frac{dr}{dt} \bigg |  = 2.7 \frac{M_\text{p}}{M_*} \frac{\Sigma r^2}{M_*} \bigg (\frac{r}{H} \bigg )^2 r \Omega,
\end{equation}
\noindent where M$_\text{p}$ is the mass of the planet, $\Omega$ is the Keplerian angular velocity and $H$ is scale height of the disk at the planet's orbital radius, $r$ \citep{Ward1997, Tanaka2002, Bate2003}. 

At larger planet masses, Type II migration occurs as the planet opens a gap in the disk. The planet then migrates at the viscous radial velocity of the disk, given by
\begin{equation}
    \label{eq:typeII}
    v_\text{II} = \bigg | \frac{dr}{dt} \bigg |  = \frac{3 \nu}{2 r},
\end{equation}
\noindent where $\nu$ is the kinematic turbulent viscosity of the disk \citep{Lin1986, Bate2003}. 

\citet{Bate2003} found that the migration rate as a function of planet mass over both regimes could be fitted by
\begin{equation}
    \label{eq:typeIandII}
    v = \bigg | \frac{dr}{dt} \bigg | = \frac{v_\text{I}}{1 + (M_p/M_t)^3} + \frac{v_\text{II}}{1 + (M_t/M_p)^3}
\end{equation}

\noindent where transition between the Type I and Type II migration occurs at a planet mass of
\begin{equation}
    \label{eq:Mt}
    M_t = 1.8 M_* \bigg (\frac{H}{r} \bigg )^3.
\end{equation}

To estimate the feasibility of the parameters required under gas--driven migration, we assumed a disk aspect ratio of $h = H/r = 0.1$ based on modelling with ALMA observations \citep{Matra2019}. The transition between Type I and Type II migration is then expected to occur at a planet mass of $M_t = 3$~M$_\text{Jupiter}$. We also assumed a surface density corresponding to the Minimum Mass Solar Nebula \citep{Hayashi1981}, in which $\Sigma = 1700 \ (r/1~\text{au})^{-3/2}$~g~cm$^{-2}$. 

Fig.~\ref{fig:migration} shows the Type I and Type II migration rate achieved at $r = 62$~au as a function of planet mass. The two curves plotted assume a viscosity of $\nu = 10^{-5}$ and $10^{-6}$ times $r^2 \Omega$, resulting in different Type II migration speeds. These values of $\nu$ are comparable to the ones used in the models by \citet{Bate2003} ($\nu = 10^{-5} r^2 \Omega$) and \citet{Crida2006} (between $\nu = 10^{-3.14}$ and $10^{-7.5}$ times $r^2 \Omega$), in which $\nu/(r^2 \Omega)$ is assumed to be constant in space and time. The two values of $\nu$ correspond to angular momentum transport efficiencies, $\alpha = \nu/(r^2 \Omega h^2)$, of $10^{-3}$ and $10^{-4}$ respectively. 

The dark--shaded region in Fig.~\ref{fig:migration} corresponds to the region of parameter space with a sufficiently fast migration speed required under the resonance model and a planet mass large enough to trap planetesimals into the 2:1(l) resonance without being too large so as to primarily trap planetesimals into the 2:1(u) resonance \citep{Wyatt2003}. This analysis therefore suggests that it is possible for gas--driven migration to provide the migration rate for the planet mass required under the resonance model.
\bsp
\label{lastpage}
\end{document}